\documentclass[journal,10pt]{IEEEtran}
\usepackage{amsmath}
\usepackage{amsfonts}
\usepackage{graphicx, dblfloatfix}
\usepackage{xcolor}
\usepackage{booktabs}
\usepackage[numbers,square,comma,compress,sort]{natbib}
\usepackage{multirow}
\usepackage{soul}

\newcommand{\gk}[1]{\textcolor{black}{#1}}

\newcommand{\rev}[1]{\textcolor{black}{#1}}

\begin{document}

\title{COIN: Communication-Aware In-Memory Acceleration for Graph Convolutional Networks}

\author{Sumit~K.~Mandal,~\IEEEmembership{Student~Member,~IEEE,} Gokul~Krishnan,~\IEEEmembership{Student~Member,~IEEE,} \\
A.~Alper~Goksoy,~\IEEEmembership{Student~Member,~IEEE,} Gopikrishnan~Ravindran~Nair,~\IEEEmembership{Student~Member,~IEEE,} \\  Yu Cao,~\IEEEmembership{Fellow,~IEEE,}~Umit~Y.~Ogras,~\IEEEmembership{Senior~Member,~IEEE}
\thanks{This work was supported in part by NSF~CAREER~Award CNS-1651624 and Semiconductor Research Corporation under task ID 3012.001, C-BRIC, one of the six centers in JUMP, a Semiconductor Research Corporation program sponsored by DARPA.}
\thanks{Sumit K. Mandal, A. Alper Goksoy and Umit Y. Ogras are with the Electrical and Compute Engineering Department, University of Wisconsin, Madison 53706, WI; Gokul Krishnan, Gopikrishnan Ravindran Nair and Yu Cao are with the School of Electrical, 
Computer and Energy Engineering, Arizona State University, Tempe 85287, AZ.
\protect \\	E-mail: \{skmandal, agoksoy, uogras\}@wisc.edu \{gkrish19, graveen1, Yu.Cao\}@asu.edu}
}

\maketitle

\begin{abstract}
Graph convolutional networks (GCNs) have shown remarkable learning capabilities when processing graph-structured data found inherently in many application areas. 
GCNs distribute the outputs of neural networks embedded in each vertex over multiple iterations to take advantage of the relations captured by the underlying graphs. Consequently, they incur a significant amount of computation and irregular communication overheads, which call for GCN-specific hardware accelerators. 
To this end, this paper presents a communication-aware in-memory computing architecture (COIN) for GCN hardware acceleration. Besides accelerating the computation using custom compute elements (CE) and in-memory computing, COIN aims at minimizing
the intra- and inter-CE communication in GCN 
operations to optimize the performance and energy efficiency.
Experimental evaluations with 
widely used datasets show up to 105$\times$ improvement in energy consumption compared to state-of-the-art GCN accelerator.
\end{abstract}

\section{Introduction}\label{sec:intro}

Graph convolutional networks (GCNs) have shown tremendous success for various applications, including node classification, social recommendations, and link predictions~\cite{duvenaud2015convolutional,ying2018hierarchical,dai2018learning}.
Their powerful learning capabilities on graphs
have attracted attention to additional research areas like image processing and job scheduling~\cite{wang2019graph,mao2019learning}.
Consequently, leading technology companies, including Google and Facebook, have developed libraries and computing systems for GCNs~\cite{lerer2019pytorch,deepmind}, stimulating further research on joint hardware and algorithm optimization.


GCNs operate on graphs by preserving their interconnections.
They have irregular data patterns since the relation between the nodes, i.e., the edge connections, do not necessarily follow a specific pattern.  
In strong contrast, classical convolutional neural networks (CNNs) are optimized for regular data patterns, which prevents them from capturing the connectivity information in the graph. GCNs use a neighbor aggregation scheme that computes each node's features using a recursive aggregation and transformation operation.
The aggregation process depends on the graph structure, while the transformation process uses a technique similar to CNN computations. 
These processes repeat until embeddings for each node are generated at the end.
As the data is sparse, irregular, and high dimensional, general-purpose platforms like CPU and GPU require energy-intensive memory accesses even if they use complex caching and prefetching techniques~\cite{chung2012application}.
Hence, the state-of-the-art GCN models are large and complex~\cite{giles1998citeseer, hamilton2017inductive, kipf2016semi}.
Multiple software-based techniques have been proposed to reduce the computation by utilizing the sparsity of the graph~\cite{tian2020pcgcn, liu2015framework}.
However, GCN execution still suffers from high latency and energy consumption.

The prevalence and computational complexity of GCNs call for high-performance and energy-efficient hardware accelerators. 
In contrast to software implementations, hardware accelerators perform GCN computations with significantly lower latency and higher energy efficiency~\cite{yan2020hygcn, geng2020awb, arka2021regraphx}. 
Due to this potential, a couple of recent studies proposed GCN accelerators~\cite{yan2020hygcn, liang2020engn}. 
These techniques implement systolic array-based architectures to perform the computations.
Since this approach requires a large number of weights, the resulting GCN hardware accelerators need a substantial number of memory accesses to fetch the weights from off-chip memory.
In turn, frequent off-chip memory accesses lead to higher latency and energy consumption as off-chip memory access consumes on average 1,000$\times$ more energy than computation~\cite{horowitz20141}.  
Therefore, there is a critical need to minimize the latency and energy consumption due to the off-chip memory accesses in GCN accelerators. 

\begin{figure}[t]
	\centering
	\includegraphics[width=1\columnwidth]{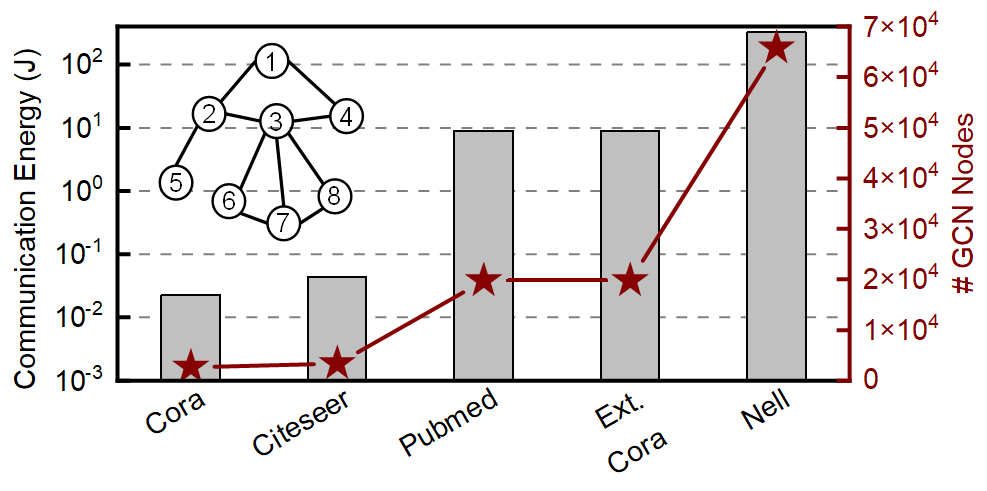}
 	\vspace{-3mm}
    \caption{Communication energy with a baseline IMC-based GCN accelerator. In the baseline architecture, the number of compute elements is equal to the number of GCN nodes and compute elements are interconnected by a 2D mesh NoC through a dedicated router. The x-axis is sorted by increasing number of GCN nodes.}
    \vspace{-3mm}
	\label{fig:energy_contri}
\end{figure}

In-memory computing (IMC) decreases
memory access-related latency and energy consumption by integrating computation with memory accesse~\cite{song2017pipelayer}.
A notable example is the crossbar-based IMC architecture, which provides a significant throughput boost for hardware acceleration by storing the weights on the chip.
However, crossbar-based in-memory computing dramatically increases the volume of on-chip communication when all weights and activations are stored on-chip.
In turn, the on-chip communication energy also increases exorbitantly. 
We implemented an IMC-based GCN accelerator baseline for popular benchmarks to quantify this effect.
Each node in the GCN is implemented using a
compute element (CE) (array of IMC crossbars) that performs the required operations. 
The CEs that make up the design are interconnected by a 2D mesh network-on-chip (NoC) through dedicated routers.
\rev{GCNs consist of thousands of nodes. The connections between the nodes enable message passing. The message passing between the nodes result in high communication volume for GCN accelerators. For example, the Nell dataset with 65755 nodes results in up to 2.7 TB of data communicated between nodes. The high volume of communication data increases communication energy consumption.}
Figure~\ref{fig:energy_contri} shows that the communication energy increases with the number of GCN nodes.
Furthermore, larger GCNs require more compute elements and routers, leading to increased chip area.
\textit{Therefore, designing an efficient on-chip communication architecture is crucial for the in-memory acceleration of GCNs.}

This paper proposes a communication-aware in-memory computing architecture (COIN) for GCN hardware acceleration. 
The COIN architecture distributes the GCN computations into multiple compute elements called CEs. 
Each CE utilizes RRAM-based crossbars for computation, significantly reducing frequent off-chip memory accesses.
Furthermore, it considers the intra- and inter-CE communications to design an optimized on-chip interconnection network.
\rev{Specifically, we construct an objective function that represents the energy consumption of communication.}
We show that the objective function is convex.
Then, we minimize the objective function to obtain the number of CEs. 
Note that the proposed methodology is also applicable to SRAM-based IMC.



The major contributions of this work are as follows:
\begin{itemize}
    \item A novel RRAM-based IMC architecture, COIN, for GCN acceleration that utilizes a communication-aware IMC architecture and a novel dataflow, 
    \item An methodology to determine the optimal number of compute elements (CEs) in COIN that ensures a balance between intra-CE and inter-CE data communication for GCN acceleration,
    \item Experimental evaluation across popular graph datasets for GCN and comparison with respect to state-of-the-art GPUs and accelerator. COIN achieves up to 105$\times$ lower energy consumption with respect to state-of-the-art GCN accelerator.
\end{itemize}

The rest of the paper is organized as follows. Section~\ref{sec:related_work} discusses the related work, while Section~\ref{sec:background} provides background on GCNs and IMC. Section~\ref{sec:methodology} and Section~\ref{sec:expt_top} present the proposed COIN architecture and its evaluation, respectively. Finally, Section~\ref{sec:conclusion} concludes the paper.
\section{Related Work}
\label{sec:related_work}


Graph processing accelerators have recently attracted attention due to their significant impact potential.
GraphR~\cite{song2018graphr} graph processing accelerator uses two components: memory and graph engine, both based on resistive random access memory. 
\rev{It performs the graph computations in matrix format without optimizing sparsity.}
The technique shows around 16$\times$ speedup compared to CPU baseline systems.
A more recent graph processing accelerator, GraphS~\cite{angizi2019graphs}, uses spin-orbit torque magnetic random access memory (SOT-MRAM) for parallel computations to accelerate graph processing applications.
It achieves around 5$\times$ speedup compared to processing depending on DRAM acceleration.
GCNs involve convolution operations in addition to graph processing.
\textit{Therefore, accelerators that target only graph processing are not suitable for GCNs.}

A few recent studies propose GCN hardware accelerator architectures~\cite{yan2020hygcn, geng2020awb}.
For example, HyGCN~\cite{yan2020hygcn} uses a hybrid system to incorporate convolution operation and tackle the irregularity of the GCN structures.
It first divides the computations into two as aggregation and combination to exploit different levels of parallelism.
Then, a task scheduler is used to exploit edge-level parallelism by sending edge processing loads onto single instruction multiple data cores.
The combination phase performs the transformation process by utilizing a systolic-array structure.
Main memory accesses take up a significant portion of the total execution time, although the HyGCN employs multiple optimizations to reduce DRAM accesses.
Similarly, the GRIP~\cite{kiningham2020grip} architecture also divides the GCN computations into aggregation and combination engines.
It employs a parallel prefetch-and-reduce engine to handle irregular data for aggregation.
Another recent technique, EnGN~\cite{liang2020engn}, uses a ring-edge-reduce-based approach for data transfer.
This approach sends the output data to the subsequent processing unit in a physical ring system for the aggregation phase.
However, most of the execution time comes from DRAM accesses as EnGN utilizes the main memory to load the weights to process.
Similarly, Rubik~\cite{chen2020rubik} uses graph reordering, mapping-aware data reuse to achieve a better graph-level data locality. It uses a customized cache design for graph-level data reuse~\cite{abadal2020computing}.
Another recent proposal, AWB-GCN, stores the adjacency matrix and the weights on off-chip memory~\cite{geng2020awb}.
The sparse matrix multiplication kernel periodically accesses the off-chip memory and performs the computation.
It requires up to 503 GBps off-chip memory bandwidth to fully utilize the hardware.
\textit{A critical drawback of prior HW accelerators is large number of off-chip memory accesses, which increase the latency and energy consumption.}

IMC-based hardware accelerators reduce off-chip memory accesses by performing computation inside the memory element.
Thus, RRAM and SRAM-based IMC accelerators have been proposed for DNNs in the literature~\cite{qiao2018atomlayer, song2017pipelayer}.
However, IMC increases on-chip data volume, which increases latency and energy due to on-chip communication~\cite{mandal2020latency, krishnan2021impact, krishnan2021siam, krishnan2020interconnect}.
The high density and complexity of GCNs make the on-chip communication for IMC-based accelerators even more critical.
Authors in~\cite{wang2020gnn, challapalle2021crossbar} proposed an IMC-based accelerator for GCN.
However, these technique does not address the issue of the on-chip communication performance of GCN accelerators.
Recently, a RRAM-based 3D NoC-enabled accelerator for GNN training ReGraphX is proposed~\cite{arka2021regraphx}. The authors show that the proposed architecture is more energy-efficient than conventional GPUs.
\rev{More detailed survey of communication-aware IMC-based accelerators can be found in~\cite{mandal2021energy, joardar2021heterogeneous}.}


To address the limitations of prior approaches, we propose a communication-aware in-memory computing-based accelerator for GCNs.
We first identify that the large data volume in GCN results in high latency and high energy consumption due to on-chip communication.
Therefore, we co-optimize the communication energy and latency.
We determine the optimal IMC architecture for GCN acceleration, COIN through this optimization.
The communication-aware interconnect, and IMC-based computing elements significantly improve overall latency and energy for GCN acceleration.
To the best of our knowledge, This is the first communication-aware in-memory computing-based GCN accelerator.

\section{Background and Motivation}
\label{sec:background}

\subsection{Graph Convolutional Networks}
\label{ssection:gcn}

\begin{figure}[t]
 	\centering
    \includegraphics[width=1\columnwidth]{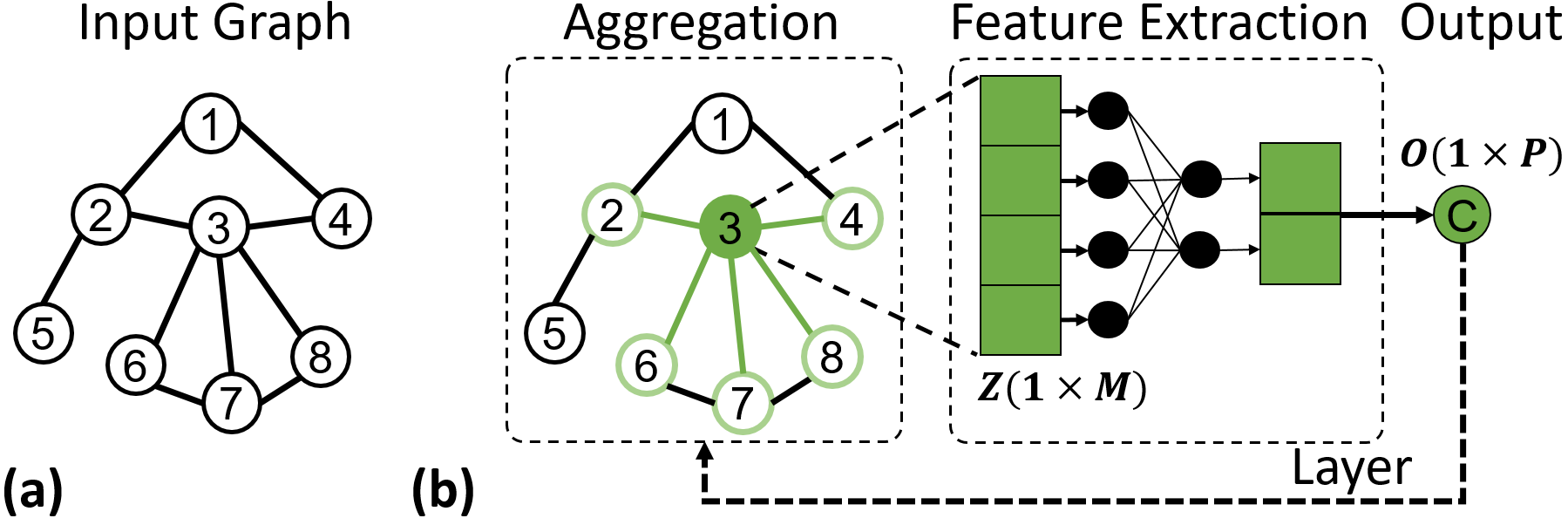}
 	\caption{(a) An example input graph, (b) Graph Convolutional Network model.} 
	\label{fig:gcn}
\end{figure}

An increasing amount of data is now represented in the form of graphs. Deep learning is effective at capturing the patterns in the Euclidean
space, but the inherent irregularity of graphs makes them unsuitable for classical deep learning techniques. 
This limitation has led to advancements in GCNs, whose structure and operations are illustrated in Figure~\ref{fig:gcn}.
GCNs maintain the graph information and can be considered as a generalized version of regular
convolutional networks.

GCN computations are divided into two stages.
First, each node aggregates the feature information from all neighbors (node 2, 4, 6, 7, 8 in Figure~\ref{fig:gcn}(a)) with its own data (node 3) during the \textit{aggregation stage}. 
For example, the $Z$ matrix in Figure~\ref{fig:gcn}(b) represents the aggregated node features from node 3 and its neighbors 
shown in Figure~\ref{fig:gcn}(a). 
As the aggregation is done by summation or averaging, the output of the aggregation stage preserves the feature dimensions,  $M\times 1$, where $M$ is the number of input features.
The aggregation stage can also consider a weighted average of neighbors' features using their node degrees~\cite{kipf2016semi}. For example, the GCN can put more weights on the neighbor nodes with lower degrees to reduce the impact of high-degree nodes.


\begin{figure*}[t]
	\centering
	\includegraphics[width=1\textwidth]{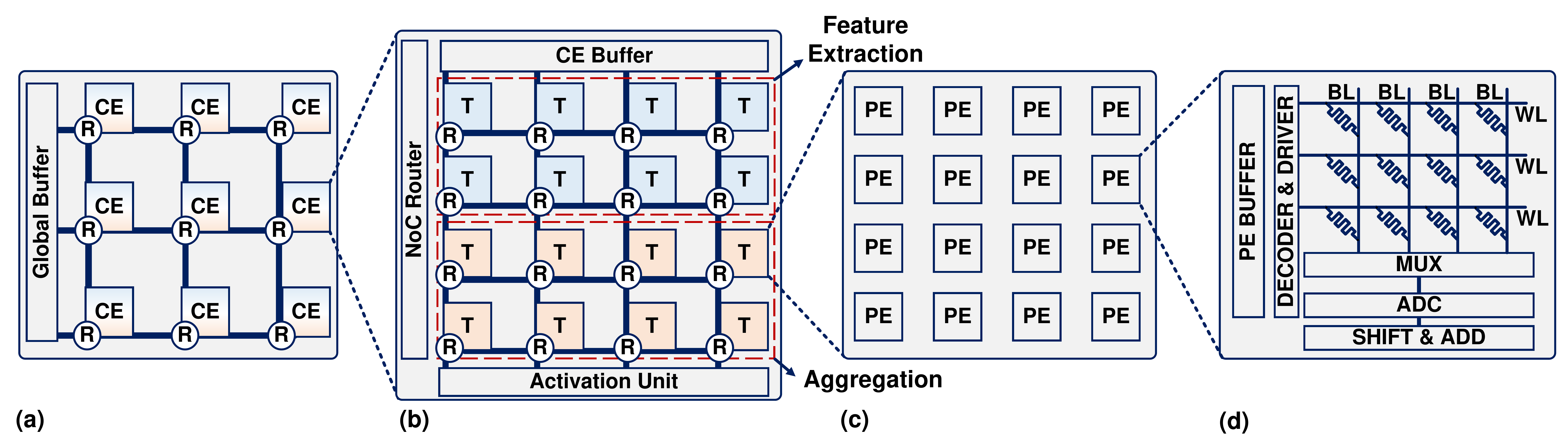}
	\caption{Overview of the COIN architecture for GCN acceleration. Each compute element (CE) consists of an array of processing elements (PEs) or RRAM-based IMC crossbar arrays connected by an NoC-mesh. A subset of the PEs performs the aggregation operation while the remaining performs the feature extraction. Aggregation PEs store the adjacency matrix while the feature extraction PEs store the layer weights.} 
	\label{fig:gnn_arch}
\end{figure*}

The second stage of GCN is the \textit{feature extraction stage}. It is similar to regular convolutional neural network computations. The result of the aggregation stage ($Z$ in Figure~\ref{fig:gcn}(b)) is fed into a multi-layer perceptron (MLP) based model. After that, an activation function like ReLU is applied. Finally, the $O$ matrix ($1\times P$, where $P$ represents the number of outputs in Figure~\ref{fig:gcn}(b)), is produced as the output of the feature extraction stage. 
These two stages repeat iteratively, where the number of layers determines the farthest distance a node feature can travel. For example, for a GCN with a single layer, each node gets information from only its neighbors. A typical GCN uses 2--3 layers~\cite{kipf2016semi}.





\vspace{-2mm}
\subsection{In-Memory Computing}
\label{section:imc}

Conventional GCN accelerators utilize separate hardware units for memory access and computation~\cite{yan2020hygcn, liang2020engn}.
Hence, GCNs with a large amount of data require a significant number of external memory accesses, which increases latency and energy consumption~\cite{horowitz20141}.
In contrast, in-memory computing combines both memory access and computation into a single unit, 
reducing the latency and energy consumption~\cite{sebastian2020memory, ielmini2018memory}.
IMC architectures utilize an array of tiles that integrates crossbars of SRAM or {RRAM}
memory cells.
In addition to the crossbar, IMC utilizes peripheral circuits, such as analog-to-digital converters (ADCs), multiplexers, switch matrix, and shift-and-add circuits to compute the output.
%
%
%
IMC-based graph convolutional networks can suffer from significant on-chip communication overhead, especially when they have large graph sizes.  For example, the NELL dataset~\cite{carlson2010toward} has 65,755 nodes, resulting in an adjacency matrix of size 65,755$\times$65,755.
The large adjacency matrix and corresponding computations incur a significant on-chip communication overhead.
Hence, there is an urgent need for an optimized IMC architecture that exploits the parallelism within the GCN operations while considering the on-chip communication overhead.




\section{The Proposed COIN Architecture} \label{sec:methodology}
\subsection{Overview of COIN Architecture}\label{ssec:arch}

Figure~\ref{fig:gnn_arch} shows the proposed COIN architecture for GCN accelerators.
It utilizes a hierarchical structure with an array of compute elements (CEs) connected by a hierarchical 2D mesh NoC. 
The global buffer in COIN is used to load the adjacency matrix, weights, and input features at the beginning of the inference.
It is connected to the CEs through the NoC-mesh interconnect.
The CE-level NoC performs the inter-layer on-chip data communication, while the local NoC within the CE performs the intra-layer communication.
The number of CEs is obtained through an optimization technique discussed in Section~\ref{sec:noc_optim}.

Each CE consists of an array of tiles, a CE buffer, activation unit (ReLU), all interconnected by the intra-CE NoC, as shown in Figure~\ref{fig:gnn_arch}(b). 
The tiles (T) consist of an array of processing elements (PEs) designed as IMC crossbar arrays. 
Each CE performs the GCN operations of a given layer and transmits the generated results to all other CEs to compute the next layer of the GCN.
The total number of tiles within the CE is limited to 30 (6$\times$5 mesh) to constrain the total chip area.
We note that this constraint does not limit the potential use of COIN for large datasets
since multiple instances of the COIN chip may be used if needed, as discussed in Section~\ref{sec:expt_top}.

GCNs perform two primary operations -- \textit{aggregation} and \textit{feature extraction}.
A subset of the tiles within each CE is 
reserved for each operation as shown in Figure~\ref{fig:gnn_arch}(b).
Aggregation tiles store the adjacency matrix in the IMC crossbar arrays, while the tiles used for feature extraction store the weights of all the layers in the GCN.
The feature extraction is performed first where matrix multiplication is performed between the input features X and the weights W to generate the extracted features Z.
The extracted features are then used to perform the aggregation operation using the tiles that map the adjacency matrix. 
\rev{The number of tiles required to map the adjacency matrix varies between 10 and 23, while the number of tiles required to map the weight matrix for the feature extraction operation varies between 3 and 7 for different GCNs.}
Therefore, both aggregation and feature extraction can be performed within each CE, thus localizing the compute for the GCN.
Consequently, the localized CE architecture inside COIN reduces the total on-chip communication data volume, carried by the inter-CE mesh NoC.
The activation unit below the bottom row implements the non-linear activation function after each layer in the GCN.
Finally, the CE buffer above the top row stores the intermediate outputs and the overall GCN layer output.

Each tile inside CE consists of an array of PEs or RRAM-based IMC crossbar arrays, as shown in Figure~\ref{fig:gnn_arch}(c).
Figure~\ref{fig:gnn_arch}(d) shows the structure of a PE within the COIN architecture.
PEs consist of an RRAM-based IMC crossbar array, a PE buffer, wordline (WL) decoder, driver circuits, and the associated read-out circuits.
The read-out circuits comprise the column multiplexers, a flash-based ADC, and a shift and adder circuit.
The IMC array utilizes analog domain computation to perform the matrix multiplication operations (aggregation and feature extraction) within the GCN.
All rows of the crossbar array are activated simultaneously to perform parallel multiply-and-accumulate operations.
Further, the read-out circuits convert the analog voltage to the digital domain.
After that, the digital outputs are transferred to the PE and CE buffers. 
Finally, the output from the CE buffer (layer output) is transmitted to all other CEs using the CE-level NoC in the COIN architecture.



\subsection{Finding the Number of Compute Elements (CEs)}
\label{sec:noc_optim}

IMC-based architecture reduces off-chip memory accesses at the expense of increased on-chip communication volume, leading to higher communication energy consumption and latency.
As illustrated in Figure~\ref{fig:energy_contri} in Section~\ref{sec:intro}, on-chip communication itself consumes 320J of energy for the Nell dataset with a baseline architecture.
Both the baseline and proposed COIN architecture perform the GCN computations in dedicated IMC compute elements. 
Furthermore, in the baseline design each IMC element is connected with a dedicated router in a 2D mesh NoC.
As a result, energy consumption and communication latency of the baseline design can be prohibitive for GCNs that process large datasets, such as the Nell dataset with 65755 nodes.
Therefore, there is a need to optimize the number of CEs to minimize the on-chip communication volume, thus the energy consumption and latency. 

This section presents an optimization technique to determine the optimal number of CEs by considering both intra- and inter-CE communication volume.
First, we show a canonical example illustrating intra- and inter-CE communication volume.
Then, we construct an objective function that captures total communication energy (both intra- and inter-CE) for a GCN.
Finally, we employ the interior point algorithm to minimize the objective function obtaining the optimal number of CEs.

\begin{figure}[b]
 	\centering
    \includegraphics[width=1\columnwidth]{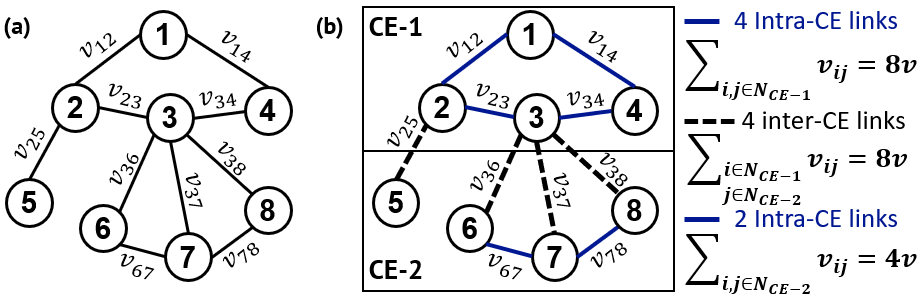}
 	\caption{A canonical graph example for intra-CE and inter-CE communication.}
	\label{fig:gcn_canon}
\end{figure}

\subsubsection{\textbf{An illustrative canonical example}}

We take a canonical example to illustrate the communication between two CEs (inter-CE) as well inside a CE (intra-CE).
Figure~\ref{fig:gcn_canon}(a) shows a graph with 8 nodes.
Each node has a convolutional network with $L$ hidden layers embedded to it.
The output of the convolutional network after each layer at each node is communicated to the nodes located at the neighbor.
Therefore, the communication between neighbors generates a communication volume. 
The bidirectional communication volume between node-$i$ to node-$j$ is represented as $v_{ij}$.

Let us assume, there are two CEs (CE-1 and CE-2) in the hardware.
Let us also assume that node 1, 2, 3, and 4 are mapped to CE-1 and node 5, 6, 7, and 8 are mapped to CE-2 as shown in Figure~\ref{fig:gcn_canon}(b).
For the sake of simplicity, we consider $v_{ij}$ are all equal ($v_{ij}=v,~\forall i,j$).
The connections inside each CE are shown in blue solid lines. 
Since there are total 4 connections inside CE-1, we obtain $8v$ as the intra-CE communication volume for CE-1.
Similarly, 2 connections inside CE-2
makes communication volume as $4v$ .
The graph in Figure~\ref{fig:gcn_canon}(b) has four 
connections between CE-1 and CE-2: 2--5, 3--6, 3--7, 3--8 shown in dashed black lines.
Therefore, the total communication volume is $8v$ between CE-1 and CE-2 (intra-CE communication).

\subsubsection{\textbf{Constructing the objective function}} 
Suppose the target GCN has $N$ nodes each implementing a convolutional network with $L$ layers, where $L,N \in \mathbb{Z}^+$.
We denote the number of input activation bits of layer $l$ as $a(l)$ for $1 \leq l \leq L$.
We note that, the communication volume between nodes ($v_{ij}$) appears due to the activation bits.
The number of output activation bits of layer $l$ can be expressed as $a(l+1) \in \mathbb{Z}^+$ since it equals to the number of input activation bits of the succeeding layer. 
Finally, let the number of CEs be denoted by the variable $k \in \mathbb{Z}^+$.
Therefore, number of nodes mapped onto each CE is $\frac{N}{k}$.
The objective function has two components which are described next.

\textit{Intra-CE Communication Energy}:
The first part of the objective function accounts for the intra-CE communication energy.
The number of CEs dictates the communication volume inside each CE.
Remember that the number of input activation bits to each node for $l^\mathrm{th}$ layer is $a(l)$.
Since we consider sparse connections between nodes, we denote the probability of having a connection between two nodes in CE-$m$ as ${p}^{(1)}_m$.
Since there are $\frac{N}{k}$ number of nodes mapped to each CE, there are total $\sum_{m=1}^{k}(\frac{N}{k})(\frac{N}{k}-1)p^{(1)}_m$ transactions between all nodes inside a CE. 
Hence, the total number of output activation bits within the CE after the $l^\mathrm{th}$ layer of operation is $\sum_{m=1}^{k}(\frac{N}{k})(\frac{N}{k}-1)p^{(1)}_ma(l+1)$.
We add the whole expression for $L-1$ layers to take account of the output activations for each layer.
Finally, assuming energy per bit is proportional to square root of number of nodes per CE, we obtain the total intra-CE communication energy as follows:

\begin{align} \label{eq:obj_func_intra}
    E_{intra-CE}(k)=\sum_{m=1}^{k} \frac{N}{k} \Big( \frac{N}{k}-1 \Big)p^{(1)}_m \sum_{l=1}^{L-1}a(l+1) \Big( \frac{N}{k} \Big)^{\frac{1}{2}}
\end{align}

\textit{Inter-CE Communication Energy}:
%
%
The second part of the objective function accounts for the inter-CE communication energy.
The number of CEs in the system is a key contributor to inter-CE communication volume.
As a reminder, each CE implements the functionality of $\frac{N}{k}$ nodes of the target GCN.
We denote the probability of having a connection between two nodes mapped in CE-$i$ and CE-$j$ as $p_{ij}^{(2)}$. 
Since each node generates $a(l+1)$ output activation bits after processing the $l^\mathrm{th}$ layer between CE-$i$ and CE-$j$, the number of output activation bits generated is $\frac{N}{k}\frac{N}{k}a(l+1)p_{ij}^{(2)}$.
We note that, all CEs generate output activation bits to $(k-1)$ other CEs.
Therefore, the total inter-CE communication volume is obtained by adding the summations to take account of all CE pairs.
Finally, assuming that the energy per bit for inter-CE communication is proportional to square root of number of CEs ($k^{\frac{1}{2}}$)~\cite{jiang2013detailed}, we obtain the total inter-CE energy by multiplying the whole expression by $k^{\frac{1}{2}}$ as follows:

\begin{align} \label{eq:obj_func_inter}
    E_{inter-CE}(k) = \sum_{i=1}^k\sum_{\substack{j=1\\j\neq i}}^k\Big(\frac{N}{k} \Big) \Big( \frac{N}{k} \Big) p_{ij}^{(2)} \Bigg(  \sum_{l=1}^{L-1}a(l+1)  \Bigg) k^{\frac{1}{2}}
\end{align}

%
%

Finally, we obtain the total communication energy by adding intra-CE (Equation~\ref{eq:obj_func_intra}) and inter-CE (Equation~\ref{eq:obj_func_inter}) communication energy as shown in Equation~\ref{eq:obj_func}.

\begin{equation} \label{eq:obj_func}
    E(k) = E_{intra-CE}(k) + E_{inter-CE}(k)
\end{equation}

\subsubsection{\textbf{Solving the objective function}}
Our goal is to minimize the objective function $E(k)$ with constraints in Equation~\ref{eq:constraints}.
As a reminder, each CE is connected to a NoC router.
Hence, the number of NoC routers is equal to the number of CEs.
The constraint in Equation~\ref{eq:constraints} states that the number of routers in the NoC ($k$) is positive and is linear on $k$.
In Appendix A, we show that $E(k)$ is convex.
Since $E(k)$ is a convex function with linear constraint, we can apply any standard algorithm to solve a convex optimization problem.
In this work, we use the interior point algorithm~\cite{karmarkar1984new} to solve Equation~\ref{eq:obj_func} with constraints in Equation~\ref{eq:constraints}.
We use this algorithm since it provides a solution in polynomial time.
Specifically, it takes only 10ms to obtain a global minimum.
\textit{Based on the result, we consider a 4$\times$4 mesh NoC to connect CEs i.e. total number of CEs in COIN is 16.}
\rev{With 16 CEs, COIN consists of 30 MB of memory on-chip.}

\begin{equation} \label{eq:constraints}
\begin{aligned}
& \underset{\mathbf{k}}{\text{minimize}}
& & E(k) \\
& \text{subject to}
& & k > 0.
\end{aligned}
\end{equation}



\subsection{Proposed Mapping and Dataflow}
\label{ssec:dataflow}


\subsubsection{\textbf{Mapping of GCN to the RRAM-based IMC crossbar arrays}}\label{sec:mapping}
This section describes the mapping of the adjacency matrix and layer weight matrix onto the RRAM-based IMC crossbar arrays. 
Since adjacency matrix remains the same for all the layers, we map the adjacency matrix onto the RRAM-based IMC crossbars inside a CE and reuse for all layers.
\rev{Specifically, if a GCN consists of $N$ nodes and the architecture has $k$ CEs, then the size of the adjacency matrix to be mapped in each CE is $N\times \frac{N}{k}$.}
We note that the weight matrices are specific to each layer and are smaller in size.
Therefore, the total number of IMC crossbar arrays required to map the weight matrices is smaller than the number of IMC crossbar arrays required to map the adjacency matrix.
We further note that the mapping of both (adjacency and weight) matrices is performed as is without any matrix transformations.

\begin{figure}[t]
	\centering
	\includegraphics[width=1\columnwidth]{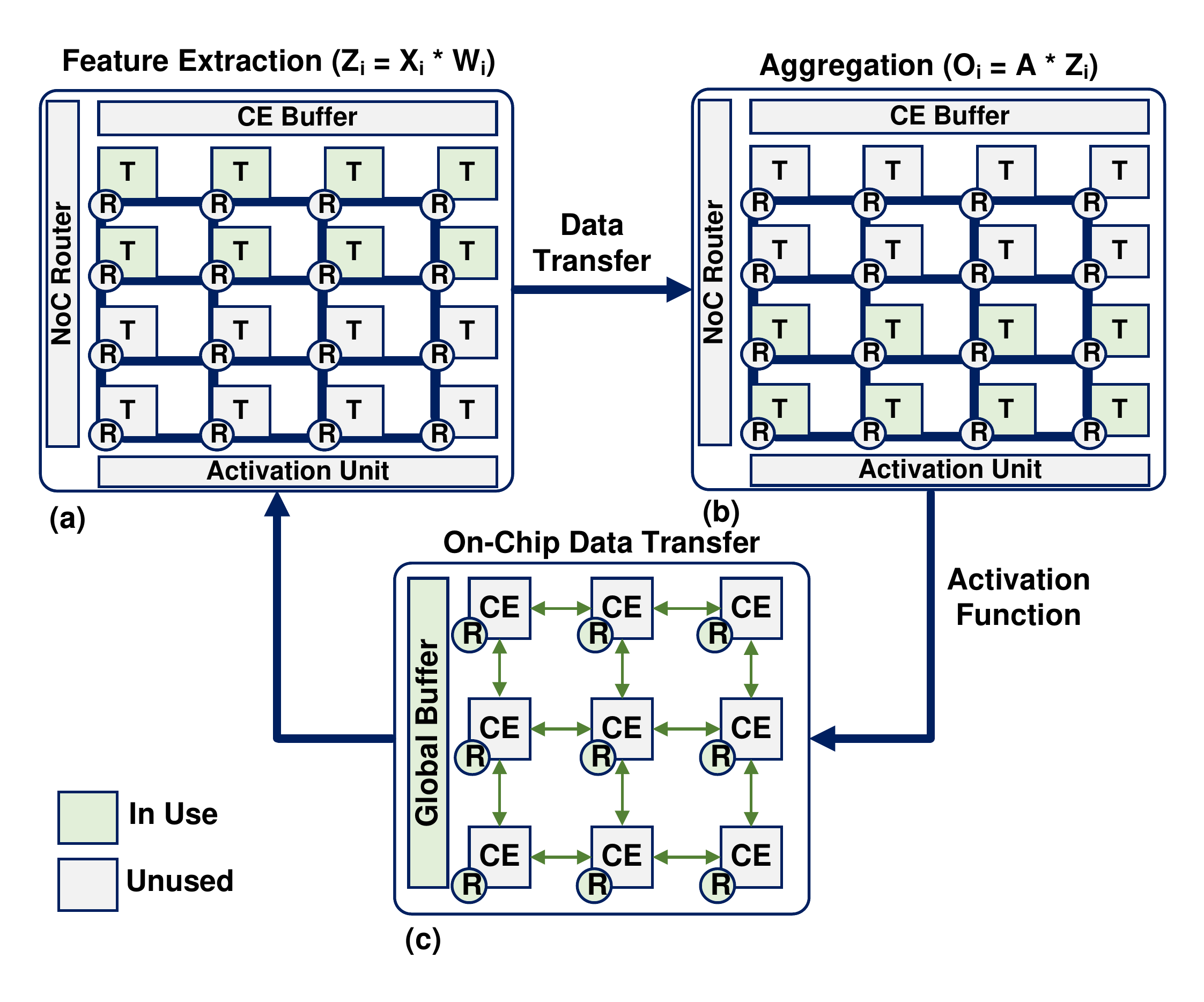}
	\caption{Layer-wise execution dataflow of the proposed COIN architecture. (a) shows feature extraction operation of layer-$i$ in a CE, (b) shows aggregation operation of layer-$i$ in a CE, (c) shows the communication between CEs.} 
	\label{fig:gnn_dataflow}
\end{figure}

\subsubsection{\textbf{Dataflow of the COIN architecture}}
We propose a layer-wise operation for the inference with COIN, as illustrated in Figure~\ref{fig:gnn_dataflow}.
In the beginning, the adjacency matrix and the weights are loaded into the corresponding IMC crossbars from the off-chip memory.
Then, the proposed layer-wise computations within each layer are performed parallely (simultaneously within each CE), while across layers are executed serially.
We denote the input features to layer-$i$ as $X_i$ and weights of layer-$i$ as $W_i$.
\gk{The feature extraction unit (IMC crossbar arrays) performs the matrix multiplication between the feature matrix ($X_i$) and the weight matrix ($W_i$) to generate the intermediate output $Z_i$ as shown in~Figure~\ref{fig:gnn_dataflow}(a).
The weights are stored in the RRAM IMC crossbars in a column-wise manner.
The transposed form of the input feature matrix $X_i$ is then provided to the crossbar array as the input vector across the wordlines.
Each input gets multiplied with the corresponding weight value stored in the RRAM to generate the output.
The computed result then accumulates along the bitline (BL) in the current domain. 
Next, the analog current is then converted to digital using an analog-to-digital converter (ADC). 
The proposed COIN architecture does not utilize a digital-to-analog converter, while it utilizes bit-serial computing for multi-bit inputs (shift and add circuit performs the accumulation based on the positional value of the input). 
}

\gk{After that, the aggregation operation is performed.
The adjacency matrix is stored in the RRAM IMC crossbar arrays to form the aggregation unit, similar to weights within the feature extraction operation.
The input of the aggregation unit is the transposed form of the intermediate output ($Z_i$) from the feature extraction operation.
The aggregation unit performs the matrix multiplication between the intermediate output $Z_i$ and the adjacency matrix ($A$) to obtain  $O_i=A.Z_i$ as shown in Figure~\ref{fig:gnn_dataflow}(b).
The computation within the aggregation unit (IMC crossbar arrays) is similar to that in the feature extraction step.
The `shift and add' unit inside PE performs the addition on the aggregated output.
Next, the ReLU operation is performed on $O_i$ to obtain the output from layer $i$ across all CEs.
Finally, the output from layer $i$ is communicated to all CEs via the NoC to perform the computation for layer $i+1$ of the GCN,
as shown in Figure~\ref{fig:gnn_dataflow}(c).}



\subsubsection{\textbf{Illustration on number of multiplication operations}}
The proposed dataflow helps to reduce the number of multiplication operations and hence the communication between feature extraction and aggregation unit within a CE.
For example, let us consider the Nell dataset and the operation of its first layer.
The size of the adjacency matrix $A$ is 65755$\times$65755, the size of the feature map is $X_1$ 65755$\times$5414, and the size of the weight matrix is $W_1$ 5414$\times$16. 
If the aggregation operation is performed first and then feature extraction, the total number multiplication operations are:  65755$\times$65755$\times$5414 (aggregation) +  65755$\times$5414$\times$16 (feature extraction) = \textbf{2.3$\times$10$^\mathrm{13}$}.
At the same time, if feature extraction is performed first and then aggregation (proposed approach),
the total number of multiplications performed is given by 65755$\times$5414$\times$16 (feature extraction) + 65755$\times$65755$\times$16 (aggregation) = \textbf{7.4$\times$10$^\mathrm{10}$}.
Therefore, there is a 311$\times$ reduction in the number of multiplication operations for Nell with our proposed dataflow for COIN.
The reduction comes from the fewer multiplication operations in the aggregation stage.

\section{Experimental Evaluation}\label{sec:expt_top}


This section present the area, latency, and energy consumption evaluations of the proposed COIN architecture.
This work assumes 32 nm process technology and 1 GHz operating frequency.

\begin{table}[t]
\centering
\caption{Properties for different GCN datasets}
\vspace{-2mm}
\begin{tabular}{|c|c|c|c|c|c|}
\hline
                 & \textbf{Cora}  & \textbf{Citeseer}  & \textbf{Pubmed} & 
                 \begin{tabular}[c]{@{}c@{}}\textbf{Ext.}\\\textbf{Cora}\end{tabular}& \textbf{Nell} \\ \hline
\begin{tabular}[c]{@{}c@{}}\textbf{Dataset} \\\textbf{Type}\end{tabular}       & \begin{tabular}[c]{@{}c@{}}Citation\\Network\end{tabular}&\begin{tabular}[c]{@{}c@{}}Citation\\Network\end{tabular}&\begin{tabular}[c]{@{}c@{}}Citation\\Network\end{tabular}&\begin{tabular}[c]{@{}c@{}}Citation\\Network\end{tabular}&   \begin{tabular}[c]{@{}c@{}}Knowled. \\ Graph\end{tabular}\\ \hline
\textbf{\# Nodes}         & 2708  & 3327  & 19717   &  19793 &  65755\\ \hline
\textbf{\# Edges}         & 10556 & 9228  & 88651   &  130622 &  266144\\ \hline
\textbf{\# Features}      & 1433  & 3703  & 500     & 8710 &  5414\\ \hline
\begin{tabular}[c]{@{}c@{}}\textbf{\# Output} \\\textbf{Labels}\end{tabular} & 7     & 6     & 3        & 70 &  210\\ \hline
\textbf{\# Layers} & 2     & 2     & 2       &  2  &  2 \\ \hline
\end{tabular}
\label{tab:dataset_config}
\vspace{-2mm}
\end{table}


\begin{table}[b]
\caption{Summary of circuit level and NoC parameters}
\begin{tabular}{|l|l|l|l|}
\hline
\multicolumn{2}{|c|}{Circuit}           & \multicolumn{2}{c|}{NoC}      \\ \hline
PE array size        & $128 \times 128$ & Bus width              & 32   \\ \hline
Cell levels          & 2 bit/cell       & Routing algorithm      & X--Y \\ \hline
Flash ADC resolution & 4 bits           & Number of router ports & 5    \\ \hline
Technology used      & RRAM            & Topology               & Mesh \\ \hline
\end{tabular}
\label{tab:circuit_param}
\end{table}
\begin{figure}[t]
	\centering
	\includegraphics[width=1\columnwidth]{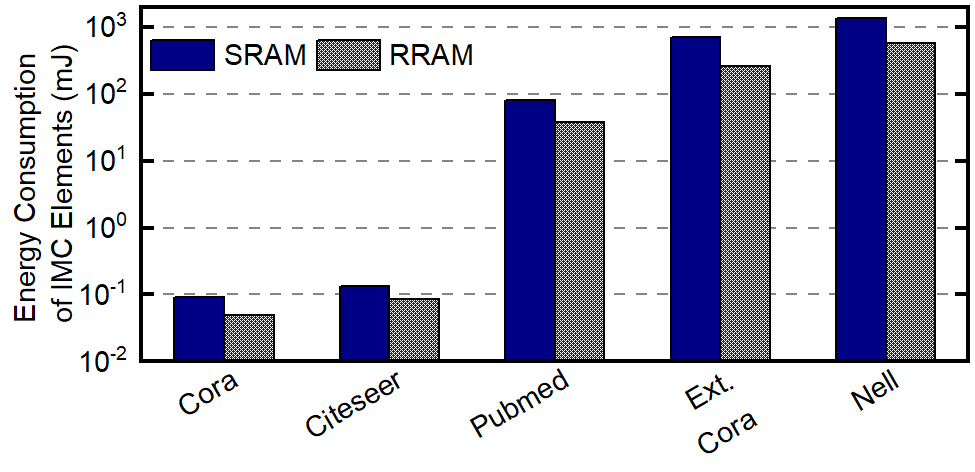}
	\caption{\rev{Comparison of Energy Consumption (in log scale) of IMC Elements between SRAM and RRAM-based designs.}} 
	\label{fig:sram_rram}
\end{figure}

\subsection{Experimental Setup}
\label{ssec:setup}

\noindent\textbf{Datasets:}
We evaluate the COIN architecture using widely used graph datasets: Cora~\cite{mccallum2000automating},  Citeseer~\cite{giles1998citeseer}, Pubmed~\cite{kipf2016semi}, Nell~\cite{carlson2010toward} and Extended Cora~\cite{bojchevski2017deep}.
Nell represents a knowledge graph dataset used to learn to read the web, and other datasets are scientific publication citation datasets for classification.
Table~\ref{tab:dataset_config} shows the details of these datasets.

\noindent\textbf{Simulation framework:}
We customized an open-source simulator~\cite{krishnan2021interconnect} to incorporate GCN attributes and support the COIN architecture. 
The inputs to the simulator include the GCN parameters, such as the number of nodes, the graph structure of each node, and input/output features for each layer. 
In addition, the simulator uses user inputs such as technology node, RRAM-based IMC crossbar size, frequency of operation, NoC size, NoC frequency, and the number of bits per RRAM cell, among others.
The performance of the computing elements is measured by customizing NeuroSim~\cite{chen2018neurosim}.
The lower-level components (e.g., buffers, ADC, multiplexers) are simulated using the Predictive Technology Model (PTM)~\cite{zhao2006new}, 
and verified against circuit simulation (e.g., SPICE),
reaching more than 90\% accuracy.
The communication performance is measured through a widely used cycle-accurate NoC simulator, BookSim~\cite{jiang2013detailed}. 
To this end, we developed a customized version of BookSim to evaluate the NoC performance that supports trace-based cycle-accurate simulation.
Since different GCNs exhibit different graph structures, we first generate traces for a given GCN.
The traces consist of the source router, destination router, and generation timestamps of each packet.
Since each layer (also known as the iteration) of the GCN is executed sequentially, we generate a separate trace file for each layer.
Then, we feed the traces to BookSim to evaluate the communication performance.
Finally, the performance of computation and communication components are combined to obtain the total performance.
Table~\ref{tab:circuit_param} summarizes the parameters used in COIN.
The simulation framework is publicly available in~\cite{coin_sim}.

\noindent\textbf{\rev{Comparison between SRAM and RRAM-based design:}}
\rev{Figure~\ref{fig:sram_rram} shows the comparison of energy consumption of IMC elements between SRAM and RRAM-based design across GCNs for different datasets. We note that the energy consumption by communication remains the same irrespective of the type of IMC elements used, since the volume of inter-CE and intra-CE communication do not change with different types of IMC elements. We observe that SRAM-based IMC elements consistently consume more energy than RRAM-based IMC elements. On-average SRAM-based IMC elements consume 2.1$\times$ more energy than RRAM-based IMC elements.
Since RRAM-based devices use analog computation, they are more energy-efficient than SRAM-based devices.
We note that the energy consumption by communication remains the same irrespective of the type of IMC elements used, since the volume of inter-CE and intra-CE communication do not change with different types of IMC element.
Therefore, we consider RRAM-based IMC elements for our architecture.
}

\subsection{Experiments with Different Quantization Bits}

\begin{figure}[t]
	\centering
	\includegraphics[width=1\columnwidth]{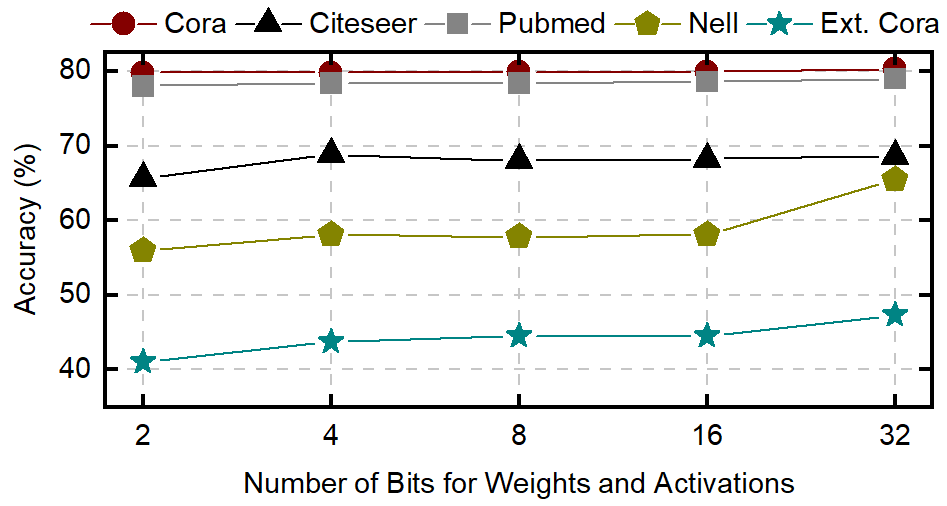}
	\caption{Accuracy with different quantization bits for weights and activations for different datasets.} 
	\label{fig:accur_bits}
\end{figure}

This section evaluates the accuracy of the GCN for different datasets with a varying number of quantization bits for weights and activations.
We used the GCN structure described by the authors in~\cite{kipf2016semi}. Deep Graph Library (DGL)~\cite{wang2019deep} with PyTorch backend and Nvidia Tesla V100 GPU are during experiments.
The accuracy for the Nell dataset increases from 55.9\% to 65.4\% when the number of quantization bits is increased from 2 to 32, as shown in Figure~\ref{fig:accur_bits}.
For Extended Cora, the accuracy varies from 41\% to 47.3\%.
For all other datasets, the difference between the minimum and maximum accuracy is less than 3\%.
In the rest of the evaluations, we consider 4-bit quantization for weights and activations since it provides comparable accuracy with 32-bit precision.

\begin{figure}[t]
	\centering
	\vspace{-3mm}
	\includegraphics[width=1\columnwidth]{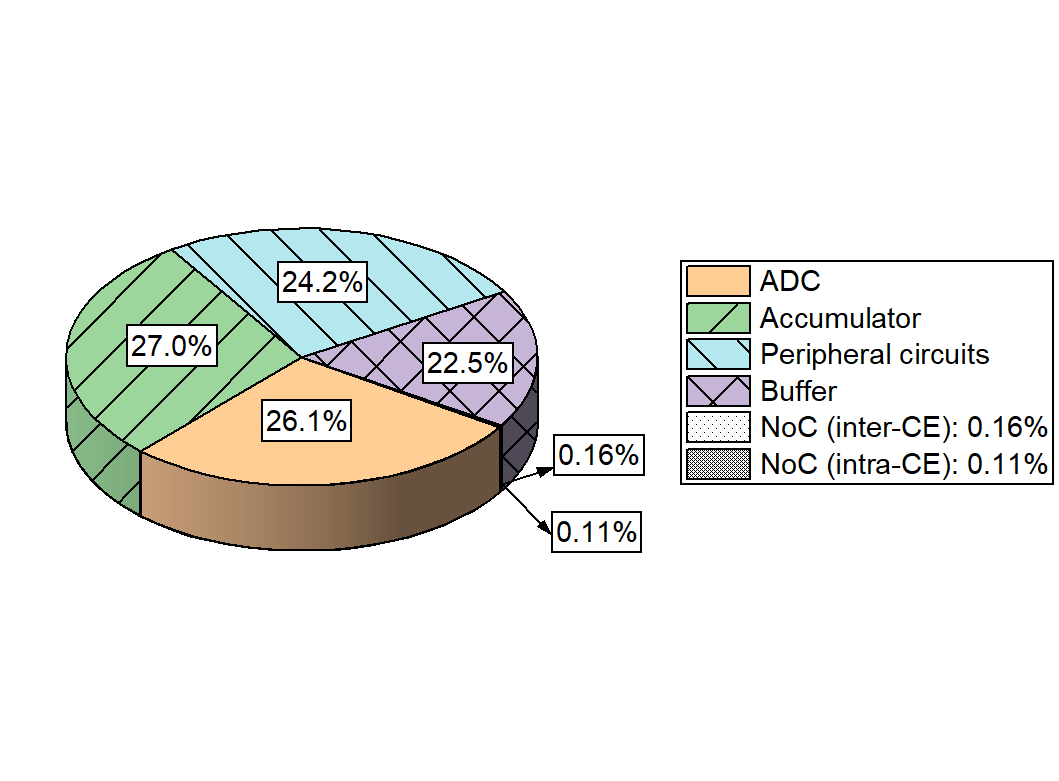}
	\vspace{-15mm}
	\caption{Different components of COIN and corresponding area.} 
	\label{fig:area_breakdown}
\end{figure}

\begin{figure*}[t]
	\centering
	\includegraphics[width=2\columnwidth]{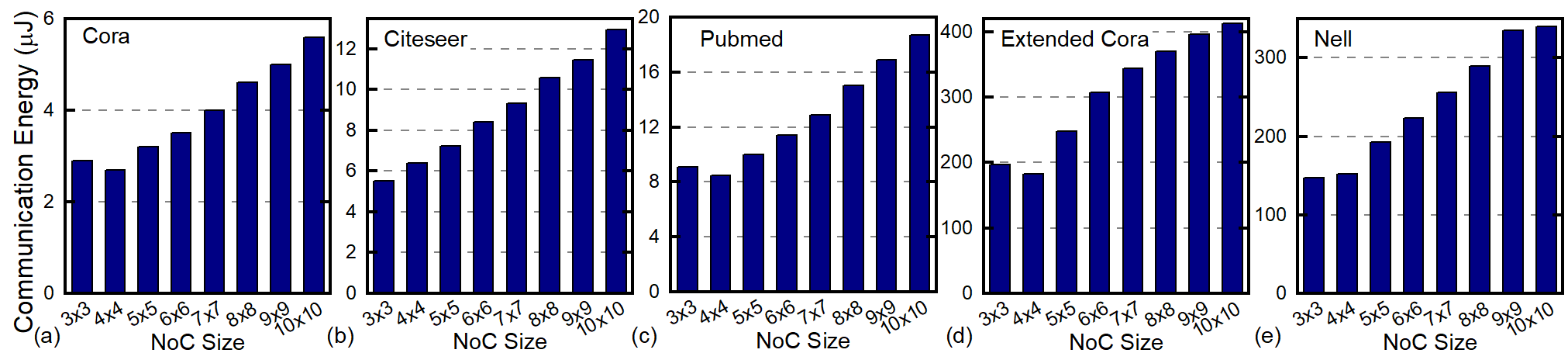}
	\caption{\rev{Comparison of communication energy consumption with different NoC sizes for (a) Cora, (b) Citeseer, (c) Pubmed, (d) Extended Cora and (e) Nell}.} 
	\label{fig:noc_perf_comp}
\end{figure*}

\subsection{Chip Area Evaluation}
The total area of COIN is 17.43 mm$^2$ with 16 CEs with 30 tiles per CE.
Figure~\ref{fig:area_breakdown} shows the area of different components of COIN responsible for computation and communication.
The components for in-memory computing include ADC to convert analog multiplication results to digital values, accumulators to perform addition operations (in Tile level), buffers to store intermediate values, and peripheral circuits.
We observe that the accumulator occupies 27\% of the total area.
The NoC for inter-CE and intra-CE communication occupy 0.16\% and 0.11\% of the total area respectively. 

We also note that GCN for large datasets such as Nell or extended Cora require multiple instances of the COIN chip.
More precisely, Cora and Citeseer require 1 chip, Pubmed requires 3 chips, extended Cora requires 20 chips, and Nell requires 45 chips.
This design choice is widely adopted for CNN accelerators (e.g. the work proposed in~\cite{shafiee2016isaac} uses up to 48 chips for a single CNN where area of each chip is 86 mm$^2$.)

\subsection{\rev{Experiments with Different Mesh Sizes}}

\rev{In this section, we compare the communication energy consumption between different NoC sizes for GCNs with different dataset.
Figure~\ref{fig:noc_perf_comp} shows the comparison.
The NoC size is varied from 3$\times$3 to 10$\times$10.
In each case, the number of CEs is equal to the number of NoC routers.
We observe that 4$\times$4 NoC (i.e. the design with 16 CEs) consumes least communication energy for most of the dataset.
For example, the least communication energy consumption for Cora dataset is 2.7 $\mu$J with 4$\times$4 NoC.
The communication energy consumption for the same dataset with 10$\times$10 NoC is 5.6 $\mu$J.
Therefore, the results with different mesh sizes show that 4$\times$4 results in the least communication energy consumption for most of the dataset which is aligned with our theoretical results.}

\subsection{Improvement with respect to Baseline} \label{sec:baseline}

\begin{figure}[t]
	\centering
	\includegraphics[width=1\columnwidth]{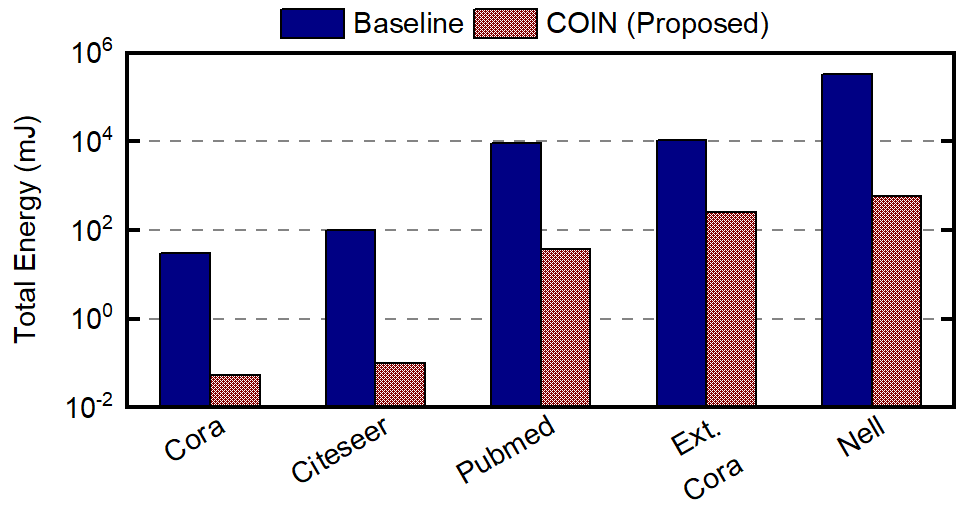}
	\caption{Comparison of total energy (in log scale) with respect to a baseline architecture. In the baseline architecture, the number of compute elements is equal to the number of GCN nodes and compute elements are interconnected by a 2D mesh NoC through a dedicated router.} 
	\label{fig:total_energy}
\end{figure}

This section compares the performance of our proposed architecture against a baseline design.
\textit{We note that a baseline design is used to show the efficacy of the proposed architecture due to the lack of prior work using IMC architectures for GCN acceleration.}
In the \textit{baseline} design, computation of each GCN node is performed using an RRAM-based IMC crossbar array, and every node is connected through a router to an NoC.
In the baseline architecture, for Cora dataset with 2708 nodes, the computations of all 2708 nodes are performed by individual IMC crossbar arrays, i.e., there are 2708 CEs.
Each CE is connected through a router of the NoC.
Figure~\ref{fig:total_energy} compares the total energy consumption of the baseline design and the proposed COIN architecture for different datasets.
We observe significant improvement in energy consumption with COIN for all datasets.
The largest improvement ($1100\times$) is obtained for the Citeseer dataset.
The GCN for the Nell dataset has the largest energy consumption (with both architectures), since Nell dataset consists of the highest number of nodes.
In this case, the baseline design consumes more than 300J energy.
However, the proposed COIN 
architecture reduces the energy consumption to 577 mJ.

We also show the percentage of the total communication energy consumption for both baseline and the proposed architecture in Table~\ref{tab:contri_energy}.
The communication energy contributes to a significant portion of the total energy with the baseline design.
For example, the communication energy makes up 96\% and 99\% of the total energy for Pubmed and Nell datasets, respectively.
With the COIN architecture, the communication energy is 7$\times$10$^{-3}$\% and 6$\times$10$^{-4}$\% of the total energy for Pubmed and Nell datasets.
Since Pubmed, Extended Cora, and Nell dataset exhibit higher sparsity than Cora and Citeseer dataset~\cite{geng2020awb}, communication energy also contributes lesser (for Pubmed, Extended Cora, and Nell) to the total energy with our proposed architecture.
The vast improvement in communication energy comes from the proposed optimization.


\begin{table}[t]
\caption{Comparison of Percentage Contribution of Communication Energy (\%).}
\centering
\begin{tabular}{|l|l|l|l|l|l|}
\hline
Datasets        & Cora & Citeseer & Pubmed & \begin{tabular}[c]{@{}l@{}}Extended\\ Cora\end{tabular} & Nell   \\ \hline
Baseline        & 43   & 44       & 96     & 58                                                      & 99     \\ \hline
COIN (Proposed) & 4.7  & 5.3      & 0.007  & 0.003                                                   & 0.0006 \\ \hline
\end{tabular}
\label{tab:contri_energy}
\end{table}


\begin{figure}[t]
	\centering
	\includegraphics[width=1\columnwidth]{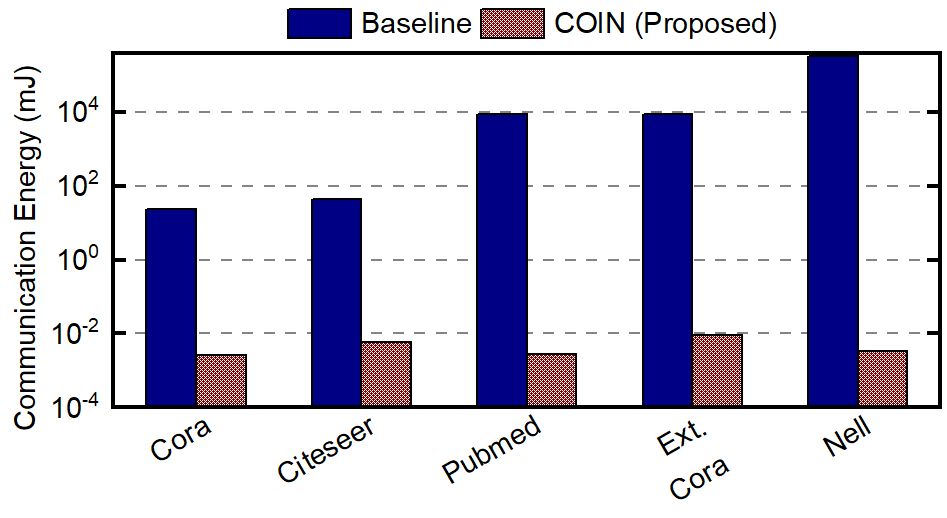}
	\caption{\rev{Comparison of communication energy (in log scale) between baseline and proposed COIN architecture.}}
	\label{fig:comm_energy_baseline}
	\includegraphics[width=1\columnwidth]{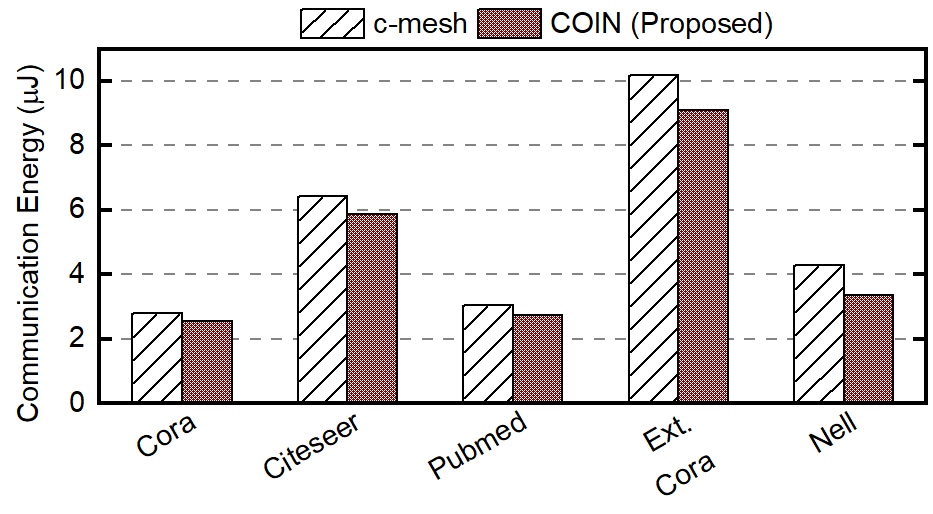}
	\caption{\rev{Comparison of inter-CE communication energy between the proposed architecture with c-mesh NoC and proposed COIN architecture with mesh NoC.}}
	\label{fig:comm_energy_cmesh}
\end{figure}

\subsection{Improvement in Communication Performance}

This section evaluates the improvement in communication performance with COIN for different datasets.
We consider the same baseline architecture as Section~\ref{sec:baseline}.
\rev{Figure~\ref{fig:comm_energy_baseline} shows the comparison of communication energy between the baseline architecture and COIN.}
Since COIN optimizes communication, there is a substantial improvement in communication energy compared to the baseline architecture.
For example, only communication itself consumes 9.2 J of energy to perform one inference in Pubmed with baseline architecture.
In contrast, COIN architecture consumes only 0.02 mJ communication energy (5 orders of magnitude improvement).
The improvement is the highest for the Nell dataset as expected (6 order of magnitude) since it has the highest number of nodes, hence the largest communication volume.
We also show the comparison of communication energy against c-mesh NoC since it has been used for accelerators targeted to CNNs~\cite{shafiee2016isaac}.
\rev{The comparison is shown in Figure~\ref{fig:comm_energy_cmesh}.}
In this case, we assume that c-mesh has 16 routers, i.e., the same number of routers as COIN. 
C-mesh uses additional links and routers, which reduces latency compared to 2D mesh.
However, c-mesh has higher energy consumption than COIN since it uses more resources.
\rev{The largest communication energy saving is observed for the Nell dataset ($1.3\times$) as shown in Figure~\ref{fig:comm_energy_cmesh}.}
Overall, COIN significantly reduces 
energy consumption compared to both the baseline architecture and c-mesh NoC.

\begin{figure}[t]
	\centering
	\includegraphics[width=1\columnwidth]{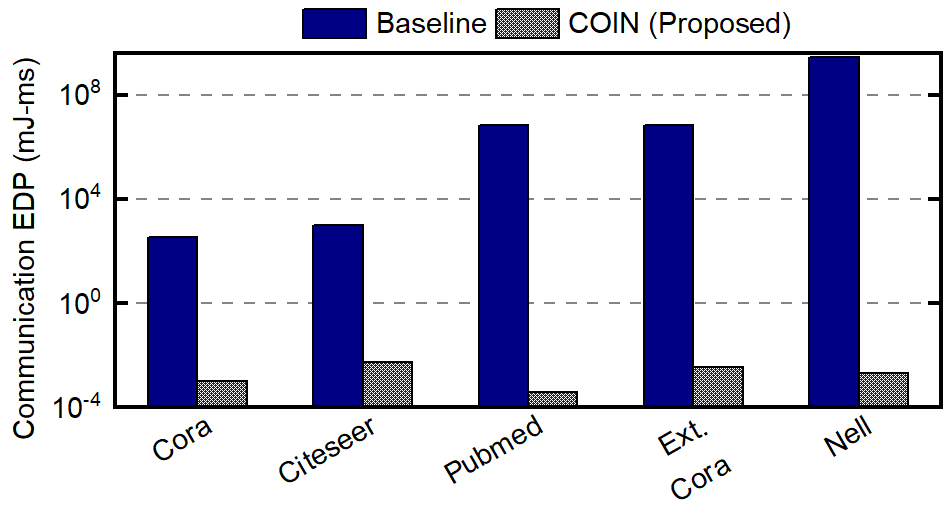}
	\caption{\rev{Comparison of EDP (in log scale) for on-chip communication between baseline and proposed COIN architecture.}} 
	\label{fig:comm_EDP_baseline}
	\includegraphics[width=1\columnwidth]{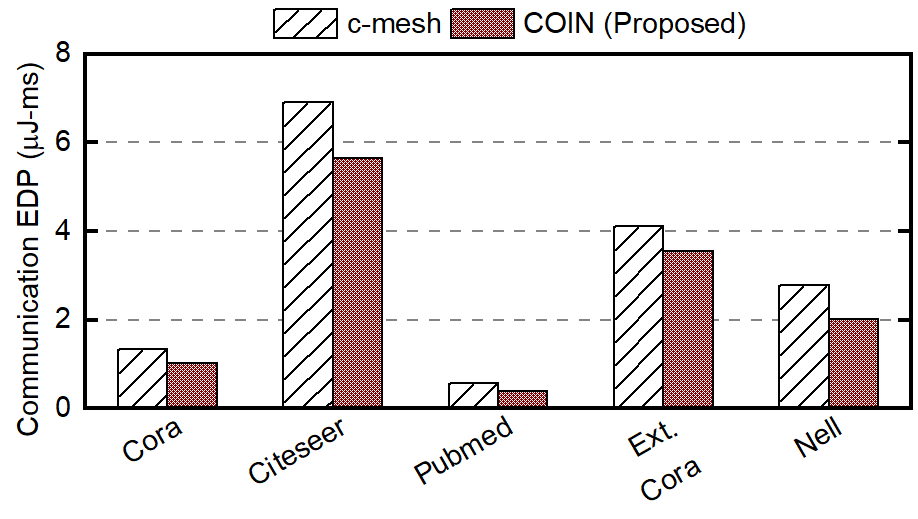}
	\caption{\rev{Comparison of EDP for inter-CE communication across baseline, proposed architecture with c-mesh NoC, and proposed COIN architecture. COIN achieves least EDP across all datasets.}} 
	\label{fig:comm_EDP_cmesh}
\end{figure}

\begin{table*}[t]
\centering
\caption{\rev{Comparison with Nvidia Quadro RTX-8000 GPU.}} \label{tab:comp_gpu}
\setlength\tabcolsep{4pt}
\begin{tabular}{|l|l|l|l|l|l|l|l|l|l|l|l|l|l|l|l|}
\hline
\multirow{2}{*}{} & \multicolumn{3}{c|}{Cora}                                          & \multicolumn{3}{c|}{Citeseer}                                      & \multicolumn{3}{c|}{Pubmed}                                        & \multicolumn{3}{c|}{Ext. Cora}                                    & \multicolumn{3}{c|}{Nell}                                           \\ \cline{2-16} 
                  & RTX  & COIN & \begin{tabular}[c]{@{}l@{}}Impr.\\ ($\times$)\end{tabular} & RTX  & COIN & \begin{tabular}[c]{@{}l@{}}Impr.\\ ($\times$)\end{tabular} & RTX  & COIN & \begin{tabular}[c]{@{}l@{}}Impr.\\ ($\times$)\end{tabular} & RTX & COIN & \begin{tabular}[c]{@{}l@{}}Impr.\\ ($\times$)\end{tabular} & RTX    & COIN & \begin{tabular}[c]{@{}l@{}}Impr.\\ ($\times$)\end{tabular} \\ \hline
Energy (mJ)      & 62.2 &     0.05  & \textbf{1244}                                                     & 90.50 &  0.10    & \textbf{905}                                                     & 89.1 &  38.13    &  \textbf{2.4}                                                    & 1787.3    &  257.4    & \textbf{6.9}                                                     & 1504 &  577.1   & \textbf{2.6}                                                    \\ \hline
Latency (ms)      & 1.22 &     0.6  & \textbf{2}                                                     & 1.22 &  1.10    & \textbf{1.1}                                                     & 1.22 &  0.57    &  \textbf{2.1}                                                    & 7.45    &  9.96    & \textbf{0.8}                                                     & 14.94 &  1.04   & \textbf{14.4}                                                    \\ \hline
EDP (mJ.ms)       & 75.78   & 0.03     & \textbf{2526}                                                      & 110.68   & 0.11     & \textbf{1006}                                                     & 108.65  & 21.56     & \textbf{5.1}                                                     & 13309  & 2564    & \textbf{5.2}                                                     & 22423  &  601.4    & \textbf{37.3}                                                    \\ \hline
\end{tabular}
\end{table*}

\begin{table}[t]
\centering
\setlength\tabcolsep{3pt}
\caption {Configuration of the edge devices considered} \label{tab:jetson_config}
\begin{tabular}{|l|c|c|c|c|c|}
\hline

           & \begin{tabular}[c]{@{}c@{}} \# CPU \\ Cores\end{tabular} & \begin{tabular}[c]{@{}c@{}}Max CPU\\ Freq. (GHz)\end{tabular} & TOPs & \begin{tabular}[c]{@{}c@{}} \# GPU Tensor \\ Cores\end{tabular} & \begin{tabular}[c]{@{}c@{}}Max GPU \\ Freq. (GHz)\end{tabular} \\ \hline
Xavier NX  & 6 & 1.4   & 21  & 48 &  1.1  \\ \hline
AGX Xavier & 8 & 2.26  & 32  & 64 &  1.37 \\ \hline
\end{tabular}
\end{table}


\rev{Figure~\ref{fig:comm_EDP_baseline} shows the comparison of energy-delay (EDP) product between the baseline and the proposed COIN architecture.
Since COIN shows significant improvement in communication energy and latency compared to the baseline design, we also observe considerable EDP savings.
For example, the Citeseer dataset shows seven orders of magnitude improvement in communication EDP compared to the baseline design.
Similar to the results for energy, we also observe improvement in communication EDP with respect to c-mesh as shown in Figure~\ref{fig:comm_EDP_cmesh}. 
The improvement compared to c-mesh is the highest for the Pubmed dataset (30\%).
In summary, our proposed optimization in NoC enables notable improvement in communication energy and EDP compared to the baseline and a design with c-mesh NoC.}

\begin{figure}[t]
	\centering
	\includegraphics[width=1\columnwidth]{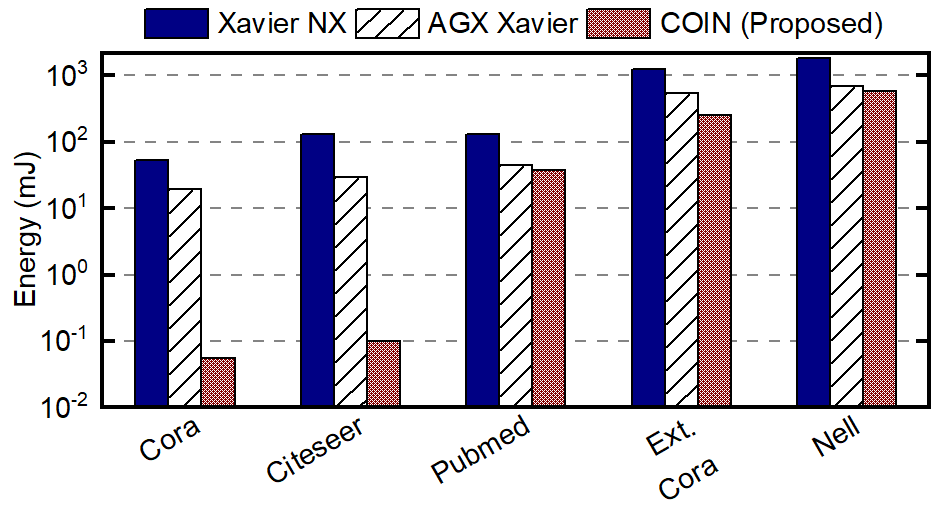}
	\caption{Comparison of energy (in log scale) between COIN and edge devices. COIN consumes less energy than both Nvidia Jetson edge devices.} 
	\label{fig:energy_gpu}	
\end{figure}

\begin{figure} [b]
    \centering
	\includegraphics[width=1\columnwidth]{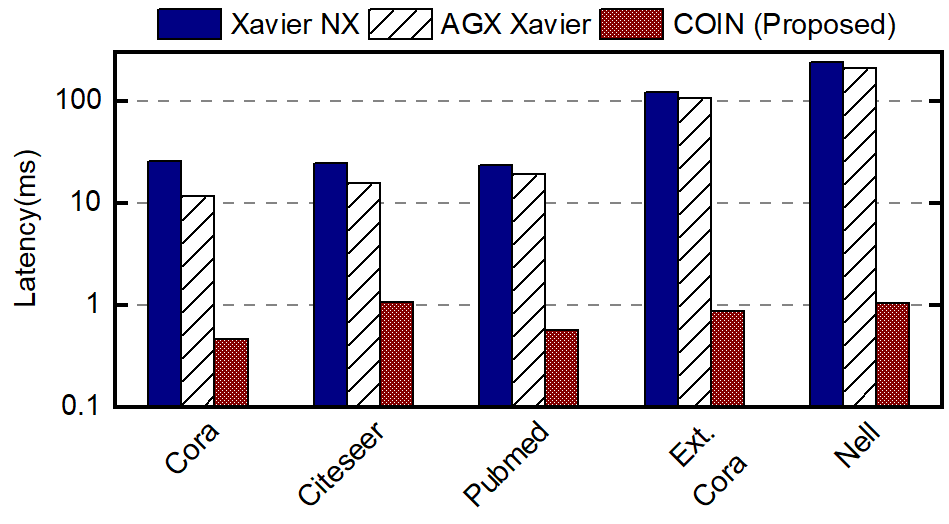}
	\caption{\rev{Comparison of latency (in log scale) between COIN and edge devices. COIN incurs less latency than both Nvidia Jetson edge devices.}} 
	\label{fig:latency_gpu}
\end{figure}

\begin{figure}[t]
	\centering
	\includegraphics[width=1\columnwidth]{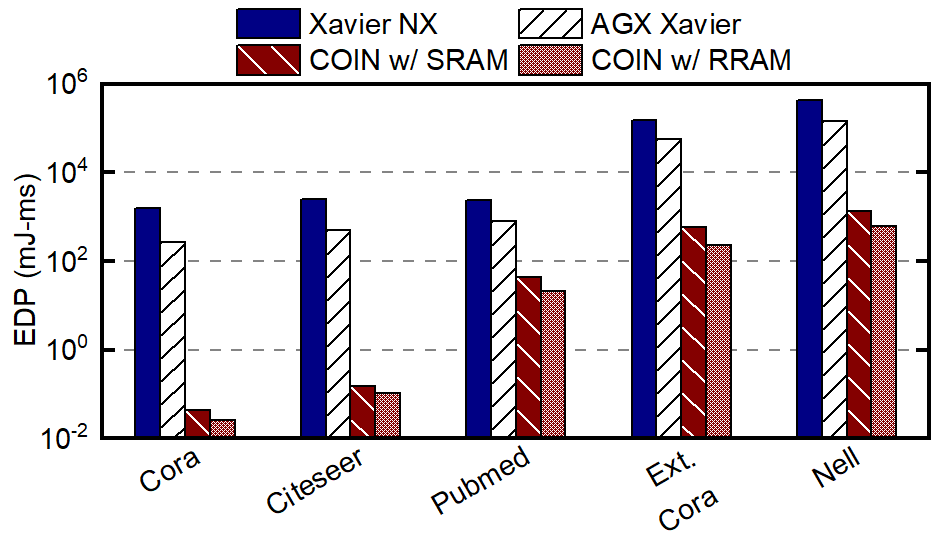}
	\caption{\rev{Comparison of EDP (in log scale) between COIN and edge devices. We present the performance of COIN with both SRAM and RRAM-based IMC elements. COIN with both kinds of devices outperforms both Xavier NX and AGX Xavier Nvidia Jetson devices across all datasets.}}
	\label{fig:edp_gpu}
\end{figure}

\subsection{Comparison with GPU and Edge Devices}

We perform a detailed comparison of the proposed COIN architecture compared to state-of-the-art GPU -- Nvidia Quadro RTX-8000~\cite{rtx8000}.
\rev{The trained GCN model for each dataset is considered, and inference is performed on the RTX-8000 GPU.}
We perform 2,000 inferences for the trained model and sample the GPU power using Nvidia's system management interface (SMI) program.
The power measurements are performed in intervals of one second.
Furthermore, we take the average of the measured power values to generate the average power for the 2,000 GCN inferences.
The inference latency is then evaluated for the GCN using python's time function.
Finally, the inference energy is evaluated by multiplying the average power and the inference latency.
We note that the same data precision of 4-bits is used for the GPU performance evaluation.

\rev{Table~\ref{tab:comp_gpu} compares the energy consumption, latency and EDP between COIN and RTX-8000.
COIN shows significant improvement both in energy consumption and EDP for all datasets compared to the GPU implementation.
For example, COIN shows 2.4$\times$ lower energy than RTX-8000 for the Pubmed dataset.
The most significant improvement in energy is observed for Cora (1244$\times$). 
Except the GCN for Extended Cora dataset, COIN shows improvement in latency over RTX-8000 GPU.
We also observe notable improvement in EDP with COIN compared to RTX-8000, as shown in Table~\ref{tab:comp_gpu}.
For Cora, COIN achieves 2526$\times$ improvement compared to the GPU.
Therefore, the proposed architecture COIN with an 
optimized NoC leads to significantly lower energy and EDP than state-of-the-art GPU.
}

We also compare the performance of our design against two edge devices - 1) NVIDIA Jetson Xavier NX and 2) NVIDIA Jetson AGX Xavier.
Such a comparison justifies the use of the COIN architecture for edge GCN inference at edge.
Table~\ref{tab:jetson_config} shows the configurations of these two devices.
We execute the GCN structures of corresponding datasets on the edge devices and record the power value at each epoch of the inference from the power sensor.
The total execution time is also recorded while executing the GCN.
Figure~\ref{fig:energy_gpu} shows energy consumption of Xavier NX, AGX Xavier, and COIN.
We observe significant improvement in energy consumption for all datasets. 
The improvement is highest for Citeseer dataset.
Specifically, for this dataset, COIN's energy consumption is 1448$\times$ and 331$\times$ lower than Xavier NX and AGX Xavier, respectively.
\rev{Figure~\ref{fig:latency_gpu} compares latency between Xavier NX, AGX Xavier and COIN across different datasets.
COIN consistently incurs less latency than both edge devices.
The highest improvement in latency is observed for Nell dataset.
COIN incurs 232$\times$ and 200$\times$ less latency than Xavier NX and AGX Xavier respectively.}
We also compare EDP between COIN and two edge devices.
A similar improvement in EDP is observed with COIN.
\rev{Figure~\ref{fig:edp_gpu} shows the comparison for EDP between COIN and edge devices. The EDP of COIN is shown considering both SRAM and RRAM-based IMC elements. On average, COIN achieves 70.7$\times$ and 50$\times$ improvement in EDP with respect to Xavier NX and AGX Xavier respectively with SRAM-based IMC elements. COIN with RRAM-based IMC elements shows 73.6$\times$ and 52.1$\times$ improvement in EDP with respect to Xavier NX and AGX Xavier respectively. The largest EDP improvement is observed for the Cora dataset with RRAM-based IMC elements. In this case, COIN achieves 4 orders of magnitude lower EDP than Xavier NX and 3 orders of magnitude lower EDP than AGX Xavier. Therefore, irrespective of the type of IMC elements used, our proposed COIN architecture achieves significantly lower energy and EDP than state-of-the-art GPU and two edge devices for a wide range of popular GCN datasets.}

\begin{figure} [b]
    \centering
	\includegraphics[width=1\columnwidth]{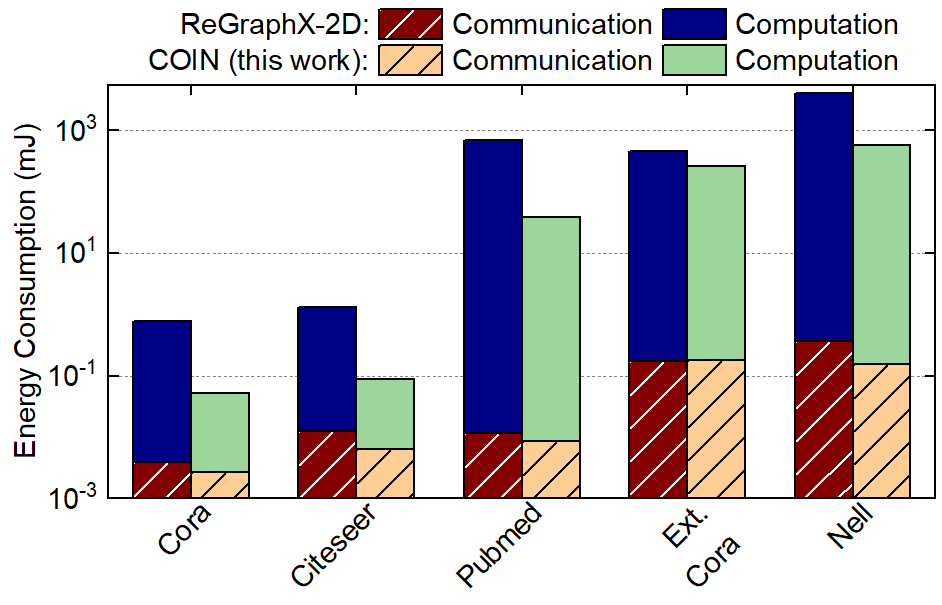}
	\caption{\rev{Comparison of energy consumption (in log scale) between 2D version of ReGraphX~\cite{arka2021regraphx} and COIN. The breakdown between communication and computation energy is shown for both the architectures.}} 
	\label{fig:regraphx_comp}
\end{figure}

\subsection{Comparison with State-of-the-art GCN Accelerators}

This section compares the performance of our proposed COIN architecture with a state-of-the-art GCN accelerator, ReGraphX~\cite{arka2021regraphx} and AWB-GCN~\cite{geng2020awb}.

\noindent{\textbf{\rev{Comparison with ReGraphX~\cite{arka2021regraphx}:}}}
\rev{The architecture proposed in ReGraphX is composed of multiple processing elements (PEs; similar to computing elements in COIN). Some of the PEs (V-PEs) store the weights and are responsible for the feature extraction operation at GCN nodes (or vertices). The other PEs (E-PEs) store the adjacency matrix of the GCN and enable ‘message passing’ through the edges of the GCN. Similarly, in ReGraphX-2D, we allocated a set of CEs for feature extraction (V-CEs) and rest to store the adjacency matrix and enable message passing (E-CEs). To have a fair comparison we consider a total of 16 CEs (same as COIN). 4 out of 16 CEs are V-CEs and 12 are E-CEs. The CEs are connected through network-on-chip routers. All CEs consist of 128$\times$128 RRAM crossbar arrays as discussed in~\cite{arka2021performance}. We consider 128$\times$128 RRAM crossbar for COIN too. For a fair comparison we evaluate the performance of ReGraphX-2D through the same simulation environment as COIN.} 

\rev{Figure~\ref{fig:regraphx_comp} shows the comparison of energy consumption between ReGraphX-2D and COIN. Specifically, we show both computation and communication components of both ReGraphX-2D and COIN. We observe that COIN consumes less energy than the ReGraphX-2D for GCN of all datasets we consider. On average, COIN consumes 8.7$\times$ less total energy than ReGraphX-2D. We also observe that communication energy consumed by COIN is consistently less than the ReGraphX-2D. For example, the GCN for CORA dataset consumes 2.7 $\mu$J and 3.9 $\mu$J of communication energy with COIN and ReGraphX-2D respectively. On average, COIN consumes 1.5$\times$ less communication energy than ReGraphX-2D across different datasets. In our proposed COIN architecture, the hierarchical communication network (inter-CE and intra-CE communication) enables more parallel communication than ReGraphX-2D. Similar to communication energy, COIN consumes less energy for computation as well compared to ReGraphX-2D. On average, the computation energy consumed by COIN is 9$\times$ less than ReGraphX-2D. In COIN, the adjacency matrix is distributed to more number of CEs compared to ReGraphx-2D. Therefore, the memory utilization of COIN is higher than ReGraphx-2D. Hence, ReGraphX-2D requires more in-memory computing (IMC) tiles than COIN which results in higher computation energy consumption with ReGraphX-2D.}

\noindent\textbf{\rev{Comparison with AWB-GCN~\cite{geng2020awb}:}}
We note that we use 32 nm technology to evaluate COIN.
However, AWB-GCN uses Intel D5005 equipped with Statix 10 SX FPGA.
This FPGA incorporates 14 nm technology.
Therefore, we estimate the performance of AWB-GCN with 32 nm technology using the technique described in~\cite{stillmaker2017scaling, sarangi2021deepscaletool}.
AWB-GCN stores the adjacency matrix and the weights in off-chip memory.
The sparse matrix multiplication kernel periodically accesses the off-chip memory and performs the computation.
Specifically, the AWB-GCN accelerator requires up to 503 Gbps off-chip memory bandwidth to fully utilize the hardware.
Since AWB-GCN uses off-chip memory, it suffers from high energy consumption.
In contrast, we use in-memory computing (IMC) to construct COIN without requiring frequent off-chip memory access.
Moreover, we incorporate an optimization technique to reduce on-chip communication energy.
The comparison in energy consumption between AWB-GCN and COIN for different GCN datasets is shown in Table~\ref{tab:sota_comp}.
\rev{We present both the energy consumption reported in~\cite{geng2020awb} and the energy consumption when the technology node is scaled to 32nm in the table.}
The energy with the Extended Cora dataset is not reported in AWB-GCN. Therefore, we cannot compare the results for Extended Cora.
\rev{The improvement is shown with respect to the energy consumption of AWB-GCN when scaled to 32nm.}
We observe that COIN provides significant improvement in energy for all the datasets we consider.
The most significant improvement is seen for the Cora dataset (105$\times$).
On average, COIN shows a 13.2$\times$ improvement in energy consumption over AWB-GCN.

\rev{Furthermore, the comparison in EDP between AWB-GCN and COIN for different GCN datasets is shown in Table~\ref{tab:sota_comp_edp}.
Similar to Table~\ref{tab:sota_comp}, we present both the EDP reported in~\cite{geng2020awb} and the EDP when the technology node is scaled to 32nm in the table.
COIN shows improvement in EDP for all the datasets we consider.
The most significant improvement is seen for the Nell dataset (7.25$\times$).
The vast improvement in energy consumption as well as EDP come from IMC-based hardware and our proposed communication-aware technique to construct the GCN accelerator.
}

\begin{table}[t]
\caption{\rev{Comparison of energy (mJ) between COIN and state-of-the-art GCN accelerator~\cite{geng2020awb}.}}
\centering
\begin{tabular}{|l|l|l|l|l|}
\hline
            & Cora & Citeseer & Pubmed & Nell \\ \hline
AWB-GCN~\cite{geng2020awb}     &    2.28  &     3.69     &      31.5    & 439     \\ \hline
AWB-GCN (scaled to 32nm)     &    5.27  &     8.54     &      73.0    & 1020     \\ \hline
COIN (ours) &    0.05  &    0.10      &     38.13    & 577.1     \\ \hline
Improvement ($\times$) &  \textbf{105}    &  \textbf{85.4}        &   \textbf{1.91}      &  \textbf{1.77}    \\ \hline
\end{tabular}
\label{tab:sota_comp}
\end{table}

\begin{table}[t]
\caption{\rev{Comparison of EDP (mJ-ms) between COIN and state-of-the-art GCN accelerator~\cite{geng2020awb}.}}
\centering
\begin{tabular}{|l|l|l|l|l|}
\hline
            & Cora & Citeseer & Pubmed & Nell \\ \hline
AWB-GCN~\cite{geng2020awb}     &    0.04  &     0.11     &      7.26    & 1425     \\ \hline
AWB-GCN (scaled to 32nm)     &    0.12  &     0.33     &      22.2    & 4358     \\ \hline
COIN (ours) &    0.03  &    0.11      &     21.58    & 601     \\ \hline
Improvement ($\times$) &  \textbf{4.68}    &  \textbf{3.09}        &   \textbf{1.03}      &  \textbf{7.25}    \\ \hline
\end{tabular}
\label{tab:sota_comp_edp}
\end{table}

\section{Conclusions}
\label{sec:conclusion}
This paper presented a novel communication-aware RRAM-based IMC architecture called COIN for GCN acceleration.
COIN utilizes an array of CEs connected through a hierarchical 2D mesh NoC optimized to balance the intra-CE and inter-CE communication volume. 
Furthermore, COIN employs CEs with an array of tiles that utilize IMC crossbar arrays.
We do not exploit the adjacency matrix and feature matrix sparsity in this work while addressing the ever-important on-chip communication cost for GCN acceleration.
The irregular structure of these matrices and the need for a column-level or block sparsity for IMC is left for further research.
Experimental evaluations across different datasets show that COIN achieves up to 105$\times$ improvement in energy consumption with respect to state-of-the-art GCN accelerator.
\rev{Other sources of energy and performance are also important in architecture exploration. However, in this work we only consider communication energy as an optimization objective.
Optimizing the other sources of energy (e.g. in-memory computing elements) are left as future work.}

\section* {Appendix A}

\noindent\textbf{Convex property of $E(k)$:} 
We show that the objective function in Equation~\ref{eq:obj_func} is convex.
To this end, we compute the second-order derivative of $E(k)$ with respect to $k$.
Equation~\ref{eq:hessian} shows the second-order derivative.
We observe that the highest probability of intra-CE connection for the dataset we consider is 0.25 and the highest probability of intra-CE connection is 0.22.
Therefore, we consider $p_{m}^{(1)}=0.25$ and $p_{ij}^{(2)}=0.22$.
%
\begin{align} \label{eq:hessian} \nonumber
   &\frac{d^2 E(k)}{dk^2} = \\
   &\Bigg(0.94\frac{N^\frac{5}{2}}{k^{\frac{7}{2}}} - 0.06\frac{N^2}{k^{\frac{3}{2}}} - \frac{0.17N^2+0.19N^{\frac{3}{2}}}{k^{\frac{5}{2}}} \Bigg) \Bigg( \sum_{l=1}^{L-1}a(l+1) \Bigg)
\end{align}
%
Equation~\ref{eq:hessian} is a function of number of routers ($k$), number of GCN nodes ($N$) and number of activations in layer-$l$ ($a(l+1)$)
We consider the range of $k$ as 4--100 to limit the NoC size as 10$\times$10.
Furthermore, we consider $N>2000$ since the smallest graph we consider is for Cora dataset with 2708 nodes.
For this range of $k$ and $N$, we observe that the first part (which is a function of $k$ and $N$ only) of the second order derivative (Equation~\ref{eq:hessian}) is always positive.
Since number of output activation bits of $l^{\mathrm{th}}$ layer is always positive, the second-order derivative, $\frac{d^2 E(k)}{dk^2}$ is positive.
As the second-order derivative of $E(k)$ is positive, $E(k)$ is convex~\cite{boyd2004convex}.

\begin{figure}[h]
	\centering
	\includegraphics[width=0.8\columnwidth]{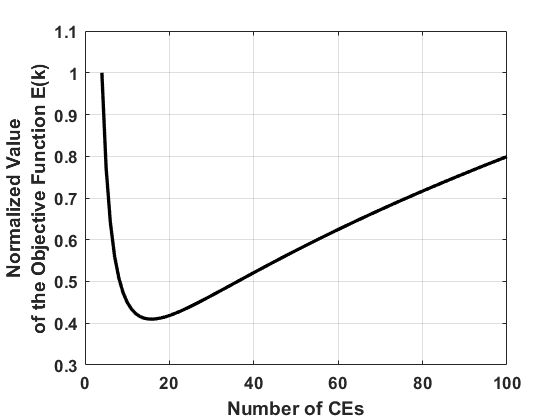}
	\caption{Normalized value of the objective function.}
	\label{fig:cvx_func}
\end{figure}

Figure~\ref{fig:cvx_func} shows the normalized value of the objective function for $N=6000$ and varying the value of $k$ from 4 to 100.
We can visualize that the objective function is convex in this case which supports our argument about the convex property of $E(k)$.

%

\small
\bibliographystyle{IEEEtran}
{\bibliography{ref}}

\begin{IEEEbiography}[{\includegraphics[width=1in,height=1.25in,clip,keepaspectratio]{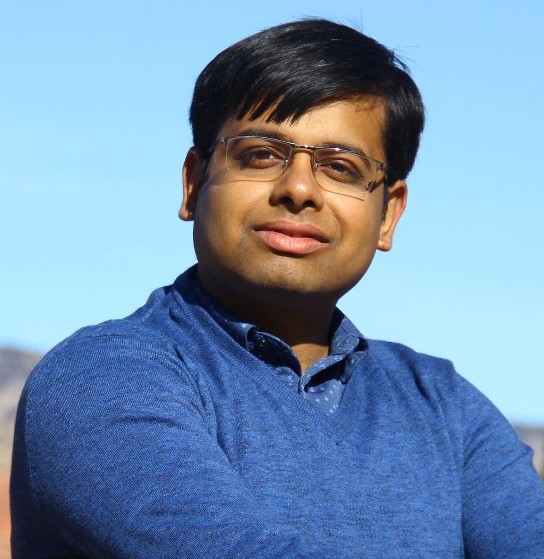}}]{Sumit K. Mandal}
(Graduate Student Member, IEEE) received the dual (B.S.+M.S.) degrees from the Indian Institute of Technology, Kharagpur, India, in 2015. He is currently pursuing the Ph.D. degree with University of Wisconsin-Madison, USA. His research interest includes analysis and design of NoC architecture, AI hardware and power management of multicore processors. He received numerous awards including best paper award from ACM TODAES in 2021.
\end{IEEEbiography}

\begin{IEEEbiography}[{\includegraphics[width=1in,height=1.25in,clip,keepaspectratio]{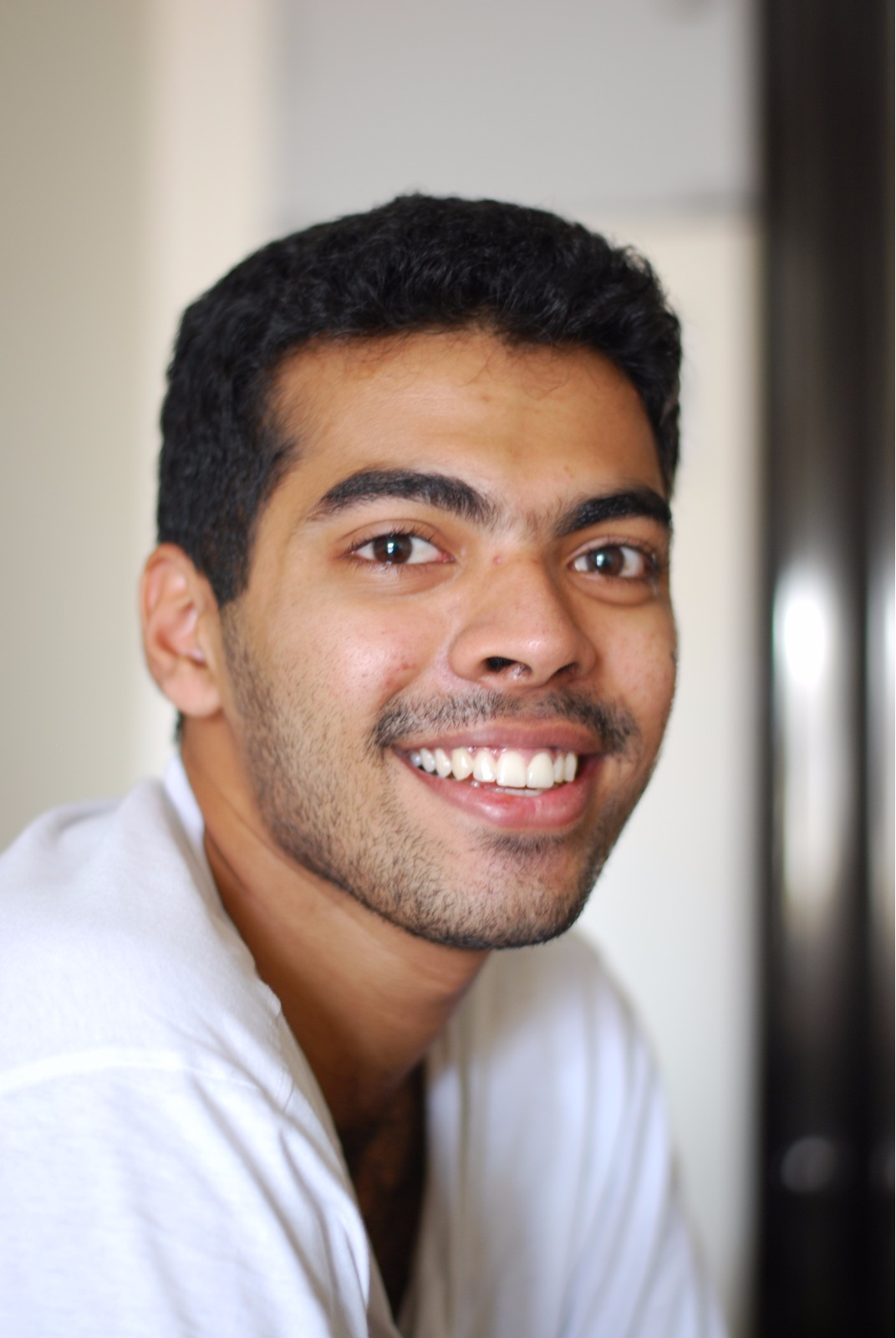}}]{Gokul Krishnan}
received his B.Tech degree in Electronics and Communication Engineering from Govt. Model Engineering College, Kochi, India in 2016. He is currently working towards a Ph.D. degree in Electrical Engineering at Arizona State University, Tempe, AZ, USA. His research interests include neuromorphic hardware and microarchitecture design for deep learning applications, chiplet-based hardware architecture for deep learning acceleration, joint algorithm-architecture design for learning on a chip, and performance modeling for CMOS and post-CMOS-based hardware architectures. He is the recipient of the Richard Newton Fellowship at DAC 2020, the Joseph A. Barkson Fellowship for 2020-21, the GPSA Outstanding Research Award in 2021, and the Engineering Graduate Fellowship in 2022. He is a student member of IEEE and a member of ACM.
\end{IEEEbiography}

\begin{IEEEbiography}[{\includegraphics[width=1in,height=1.25in,clip,keepaspectratio]{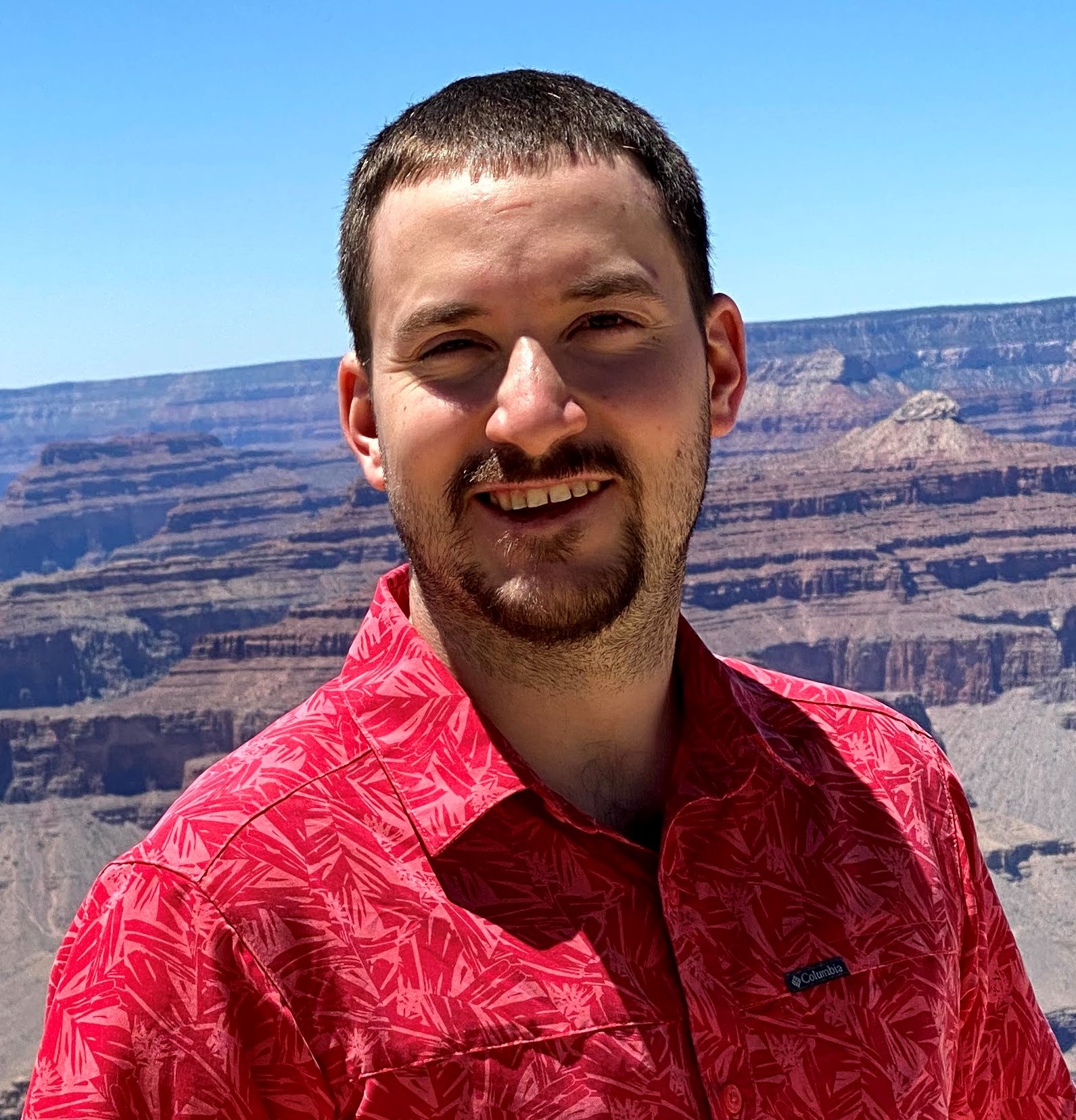}}]{A. Alper Goksoy}
received his B.S. degree in Electrical and Electronics Engineering from Bogazici University, Istanbul, Turkey.  He is currently pursuing his Ph.D. in Electrical Engineering at University of Wisconsin-Madison, USA. His research interests include task scheduling for heterogeneous SoCs, domain-specific SoCs, graph neural networks, and design of AI hardware accelerators. He received Richard Newton Young Fellowship at DAC 2019, and DAC Young Fellowship in 2021.
\end{IEEEbiography}

\begin{IEEEbiography}[{\includegraphics[width=1in,height=1.25in,clip,keepaspectratio]{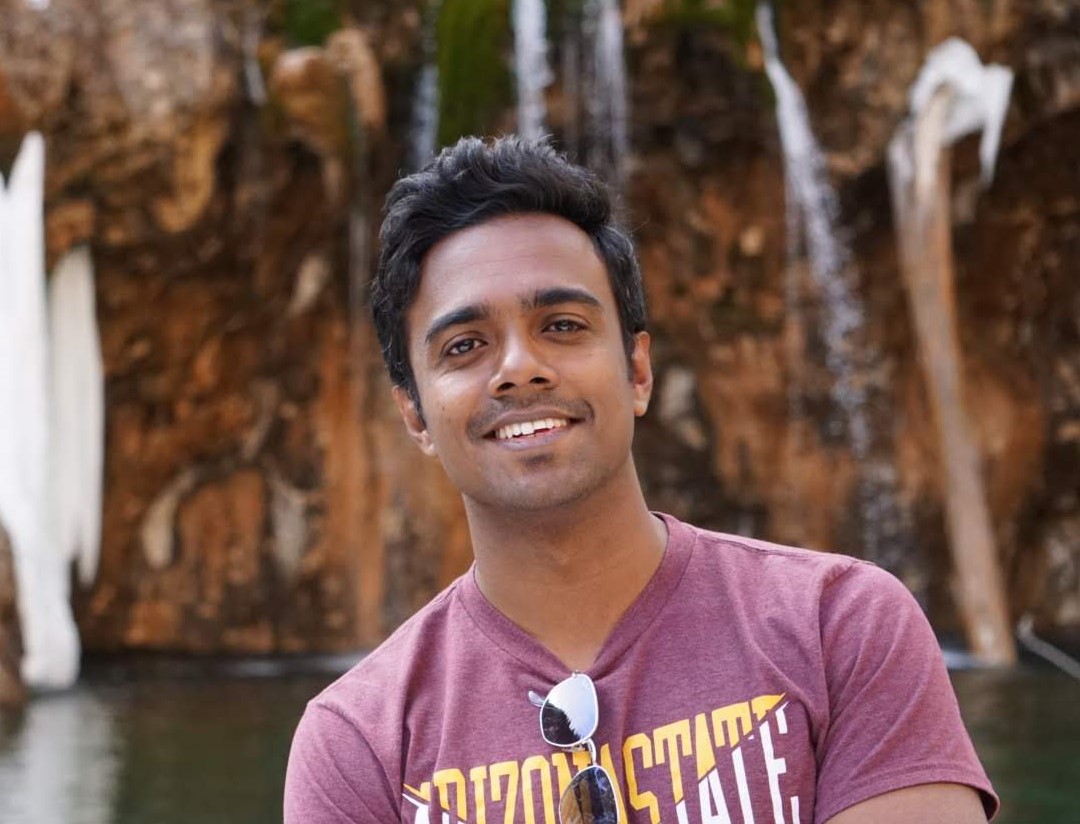}}]{Gopikrishnan Raveendran Nair} received his B.Tech in Electronics and Communications from University of Kerala, Trivandrum, India in 2014. From 2014 to 2018 he worked as a RTL design engineer at Netrasemi, Kerala, India. He completed his masters in Computer Engineering from Arizona State University in 2020 and is currently pursing his Ph.D. in Computer Engineering at Arizona State University, Tempe, USA. His research interests include hardware software co-design for machine learning applications on ASICs and FPGAs. He is a recipient of Fultion Engineering graduate fellowship in 2019 and 2021. He is a student member of IEEE.

\end{IEEEbiography}

\begin{IEEEbiography}[{\includegraphics[width=1in,height=1.25in,clip,keepaspectratio]{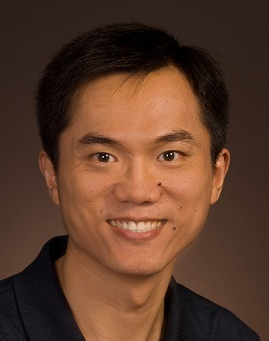}}]{Yu Cao}
(S’99-M’02-SM’09-F’17) received the B.S. degree in physics from Peking University in 1996. He received the M.A. degree in biophysics and the Ph.D. degree in electrical engineering from University of California, Berkeley, in 1999 and 2002, respectively. 
He worked as a summer intern at Hewlett-Packard Labs, Palo Alto, CA in 2000, and at IBM Microelectronics Division, East Fishkill, NY, in 2001. After working as a post-doctoral researcher at the Berkeley Wireless Research Center (BWRC), he is now a Professor of Electrical Engineering at Arizona State University, Tempe, Arizona. He has published numerous articles and two books on nano-CMOS modeling and physical design. His research interests include neural-inspired computing, hardware design for on-chip learning, and reliable integration of nanoelectronics.   
Dr. Cao was a recipient of the 2022 Best IP Award at Design, Automation and Teste in Europe Conference, the 2022 Distinguished Lecturer of IEEE Circuits and Systems Society, the 2020 Intel Outstanding Researcher Award, the 2012 Best Paper Award at IEEE Computer Society Annual Symposium on VLSI, the 2010, 2012, 2013, 2015, 2016 and 2021 Top 5\% Teaching Award, Schools of Engineering, Arizona State University, 2009 ACM SIGDA Outstanding New Faculty Award, 2009 Promotion and Tenure Faculty Exemplar, Arizona State University, 2009 Distinguished Lecturer of IEEE Circuits and Systems Society, 2008 Chunhui Award for outstanding oversea Chinese scholars, the 2007 Best Paper Award at International Symposium on Low Power Electronics and Design, the 2006 NSF CAREER Award, the 2006 and 2007 IBM Faculty Award, the 2004 Best Paper Award at International Symposium on Quality Electronic Design, and the 2000 Beatrice Winner Award at International Solid-State Circuits Conference. He served as Associate Editor of the IEEE Transactions on CAD, and on the technical program committee of many conferences.

\end{IEEEbiography}

\begin{IEEEbiography}[{\includegraphics[width=1in,height=1.25in,clip,keepaspectratio]{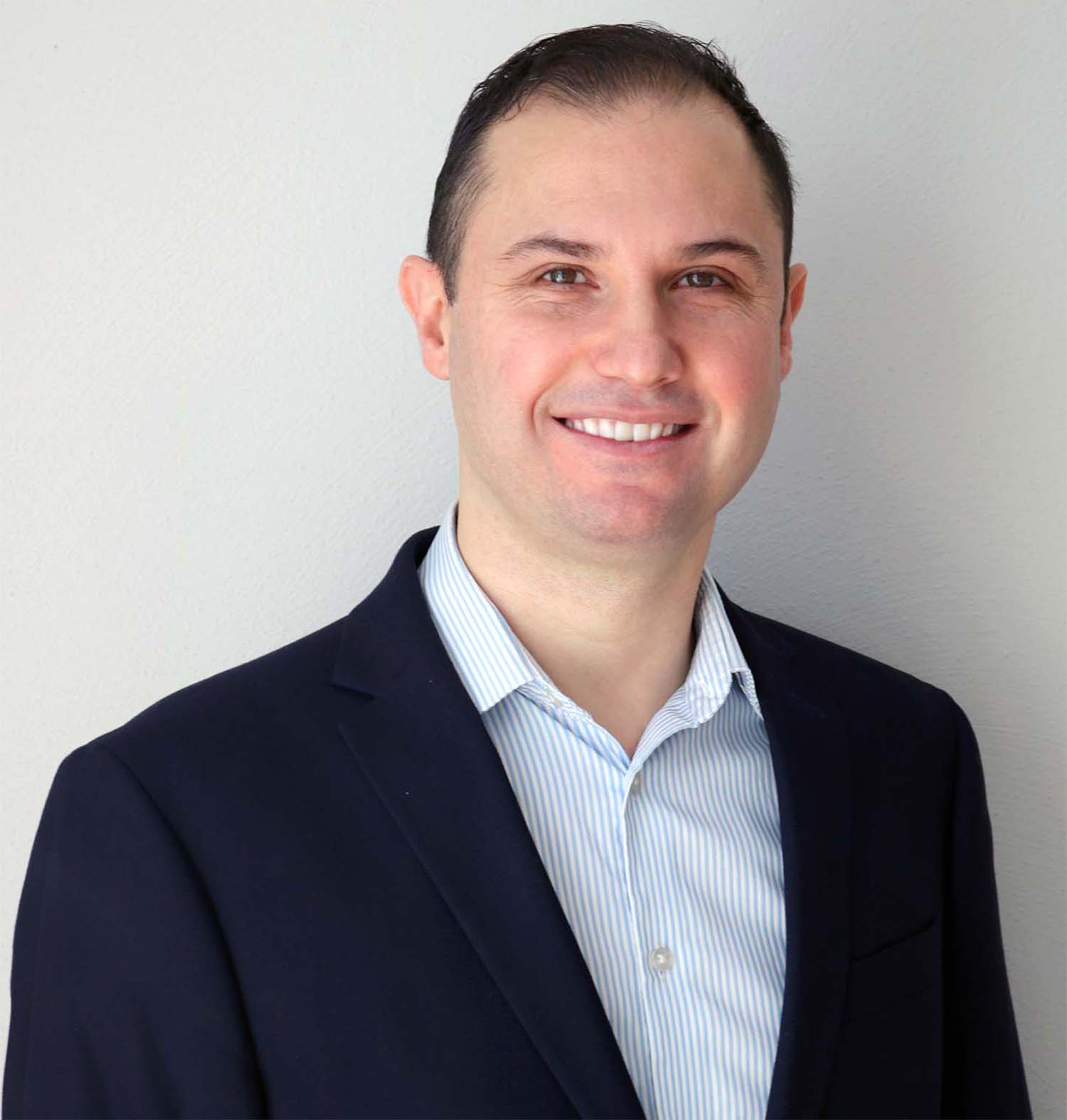}}]{Umit Y. Ogras}
received his Ph.D. degree in Electrical and Computer Engineering from Carnegie Mellon University, Pittsburgh, PA, in 2007. He is currently an Associate Professor in the Dept. of Electrical and Computer Engineering at the University of Wisconsin-Madison. His research interests include embedded systems, heterogeneous systems-on-chip, low-power VLSI, wearable computing, and flexible hybrid electronics. Dr. Ogras received DARPA Director's Fellowship Award (2020), DARPA Young Faculty Award (2018), NSF CAREER Award (2017), Intel Strategic CAD Lab Research Award (2013), and best paper awards at 2019 CASES, 2017 CODES+ISSS, 2012 IEEE Transactions on CAD, and 2011 IEEE VLSI Transactions.
\end{IEEEbiography}


\end{document}